\documentstyle[12pt,aasms4]{article}
\def\la{\mbox{\raisebox{-0.1ex}{$\scriptscriptstyle \stackrel{<}{\sim}$\,}}}
\def\ga{\mbox{\raisebox{-0.1ex}{$\scriptscriptstyle \stackrel{>}{\sim}$\,}}}

\newcommand{\Fig}{\mbox{\sc Fig. }}
\newcommand{\dmu}{ $ {\rm pc ~ cm ^ {-3} } $ }
\newcommand{\velu}{\mbox{${\rm km ~ secs ^{-1}}$\,}}
\newcommand{\nep}{\mbox{ $ {\rm N_{ep} } $ }}
\newcommand{\nsess}{\mbox{ $ {\rm N_{sess} } $ }}
\newcommand{\nref}{\mbox{ $ {\rm N_{ref} } $ }}
\newcommand{\tref}{\mbox{ ${\rm \tau _{ref}}$ }}
\newcommand{\viss}{\mbox{ ${\rm V_{iss}} ~ $ }}

\newcommand{\virr}{\mbox{${\rm V_{irr}} ~ $ }}
\newcommand{\vprop}{\mbox{ ${\rm V_{prop}}$\,}}
\newcommand{\vobs}{\mbox{${\rm V_{obs}}$ }}
\newcommand{\vu}{\mbox{ ${\rm V_{\mu}}$ }}

\newcommand{\vep}{\mbox{ $ V_{obs _{\bot }} $ }}
\newcommand{\avep}{\mbox{ $ \langle V_{obs_{\bot}} \rangle ~ $ }}
\newcommand{\cn}{\mbox{ ${\rm C_n^2}$ }}
\newcommand{\avcn}{\mbox{$ { \rm \overline { C_n^2 } } $ }}
\newcommand{\intcn}{\mbox{ ${\rm \int C_n^2}$ }}
\newcommand{\tsp}{\mbox { $ {\rm T_{sp} } $ }}
\newcommand{\tspmax}{\mbox { $ {\rm T_{sp,max} } $ }}
\newcommand{\fobs}{\mbox{ $ {\rm f_{obs} } $ }}

\newcommand{\bobsi}{\mbox{ $ {\rm B_{obs,i} } $ }}

\newcommand{\tobsi}{\mbox{${\rm t_{obs,i} } $\,}}
\newcommand{\Sort}{\mbox{ $ {\rm S_{327} } $ }}
\newcommand{\Slit}{\mbox{ $ {\rm S_{400} } $ }}
\newcommand{\avSort}{\mbox{ $ {\rm \langle S_{327} \rangle } $ }}
\newcommand{\avSlit}{\mbox{ $ {\rm \langle S_{400} \rangle } $ }}

\newcommand{\refr}{\mbox{ $ \theta _r $ }}

\newcommand{\diff}{\mbox{ $ \theta _s $ }}
\newcommand{\nd}{\mbox{$ \nu _d $ }}
\newcommand{\td}{\mbox{$ \tau _d $ }}

\newcommand{\dnt}{\mbox{ $ d \nu / d t $ }}

\newcommand{\Av}{\mbox{${\rm A_V }$ }}
\newcommand{\so}{{${\rm s_o }$}}
\newcommand{\rf}{{${\rm s_f }$}}

\newcommand{\smod}{\mbox{ ${\sigma _{mod}}$ }}
\newcommand{\sest}{\mbox{ ${\sigma _{est}}$ }}
\newcommand{\sestg}{\mbox{ ${\sigma _{est,g}}$ }}
\newcommand{\scal}{\mbox{ ${\sigma _{cal}}$ }}
\newcommand{\smodg}{\mbox{ ${\sigma _{mod,g}}$ }}
\newcommand{\smeas}{\mbox{ ${\sigma _{meas}}$ }}
\newcommand{\smeasg}{\mbox{ ${\sigma _{meas,g}}$ }}
\newcommand{\smeasf}{\mbox{ ${\sigma _{meas,F}}$ }}
\newcommand{\sfi}{\mbox{ ${\sigma _{F,i}}$ }}
\newcommand{\xrms}{\mbox{$x_{rms}$ }}
\newcommand{\sstat}{\mbox{ ${\sigma _{stat}}$ }}
\newcommand{\vmeas}{\mbox{ ${\sigma ^2 _{meas}}$ }}
\newcommand{\vstat}{\mbox{ ${\sigma ^2 _{stat}}$ }}

\newcommand{\avnd}{\mbox{ $  \langle  \nu _d \rangle  $ }}
\newcommand{\avtd}{\mbox{ $  \langle  \tau _d \rangle  $ }}
\newcommand{\avfd}{\mbox{ $  \langle  \rm F \rangle  $ }}

\newcommand{\avdtn}{\mbox{ $  \langle  d t / d \nu \rangle  $ }}
\newcommand{\ndg}{\mbox{ $ \nu _{d,g} $ }}
\newcommand{\tdg}{\mbox{ $ \tau _{d,g} $ }}

\newcommand{\dtng}{\mbox{ $ ( d t / d \nu ) _g $ }}
\newcommand{\dtn}{\mbox{ $ d t / d \nu $ }}

\newcommand{\normb}{\mbox{$ {\rm PSR ~ B0823+26(II) } $  }}
\newcommand{\egta}{\mbox{$ {\rm PSR ~ B0834+06(I) } $  }}

\newcommand{\eleva}{\mbox{$ {\rm PSR ~ B1133+16(I) } $  }}
\newcommand{\elevb}{\mbox{$ {\rm PSR ~ B1133+16(II) } $  }}
\newcommand{\elevc}{\mbox{$ {\rm PSR ~ B1133+16(III) } $  }}
\newcommand{\ninea}{\mbox{$ {\rm PSR ~ B1919+21(I) } $  }}

\begin{document}


\title{\LARGE\bf Long-Term Scintillation Studies of Pulsars: \\ I. Observations and Basic Results}

\vspace{5.0cm}

\author{\normalsize\bf N. D. Ramesh Bhat\footnote{send preprint requests to $ bhatnd@ncra.tifr.res.in $}, 
A. Pramesh Rao, and Yashwant Gupta}

\begin{center}
{\normalsize National Centre for Radio Astrophysics, Tata Institute of Fundamental Research, \\
Post Bag 3, Ganeshkhind, Pune - 411 007, India}
\end{center}

\vspace{5.0cm}

\begin{center} 
{\normalsize\bf Accepted for publication in The Astrophysical Journal Supp. Series} 
\end{center}


\begin{abstract}

We report long-term scintillation observations of 18 pulsars in the dispersion measure range $3-35$ \dmu 
carried out from 1993 January to 1995 August using the Ooty Radio Telescope at 327 MHz. 
These observations were made with the aim of studying refractive effects in pulsar scintillation,
and obtaining reliable estimates of diffractive and refractive scintillation properties.
Dynamic scintillation spectra of pulsars were regularly monitored at $10-90$ epochs spanning $100-1000$ days.
Significant changes are observed in the dynamic spectra over time scales as short as a few days.
Large-amplitude fluctuations are observed in quantities such as decorrelation bandwidth, scintillation 
time scale, drift rate, and flux density.
Several pulsars show organized features such as drifting bands in a highly pronounced manner.
For some pulsars, gradual and systematic variations are seen in the drift rate of patterns which undergo 
several sign reversals during the observing time spans.
Anomalous behaviour such as persistent drifts lasting over many months are seen for PSRs B0834+06 and 
B1919+21.
Four pulsars were studied for $2-4$ well separated observing sessions, each lasting over $ \sim $ 100 days.
In some cases, significant variations are seen in the average scintillation properties and/or flux densities 
between successive observing sessions.
From our data, we have been able to obtain more accurate and reliable estimates of scintillation properties 
and flux densities than those from the earlier observations, by averaging out the fluctuations due to 
refractive scintillation effects.
These measurements are used to derive parameters such as the strength of scattering 
and scintillation speeds.
The scintillation speed estimates are found to be reasonably good indicators of 
proper motion speeds of pulsars.
The present measurements are compared with earlier measurements and the long-term
stability of scintillation properties and flux densities is discussed.

\end{abstract}

{\it Subject headings: }{ISM:Structure -- Pulsars:General -- ISM:General -- Scattering}


\section{Introduction }

There are several classes of pulsar intensity variations seen at radio wavelengths.
The large-amplitude, broadband, pulse-to-pulse variations seen for most pulsars are 
thought to be intrinsic to the pulsar emission mechanism.
When averaged over many pulses to smooth out these variations, pulsar intensities 
show fluctuations occurring over time scales ranging from minutes to hours, which are 
explained in terms of propagation of radio waves through the irregular interstellar 
plasma (Scheuer 1968; Rickett 1969).
Random variations of electron densities in the interstellar 
medium (ISM) give rise to phase perturbations, leading to scattering of radio waves.
As they propagate, the scattered waves interfere with each other  causing large 
variations of amplitude with frequency and position.
The relative motion between the pulsar, observer and the density irregularities
translates the spatial amplitude variations into temporal variations at a given position, 
leading to  a typical time scale of intensity fading $-$ called the scintillation time scale.
Other observable consequences of this phenomenon are broadening of pulse profiles 
and angular broadening of compact radio sources at low frequencies.
Observations of pulsar dynamic spectra (e.g. Roberts \& Ables 1982; Smith \& Wright 1985), 
which are records of intensity variations 
in the time-frequency plane, reveal that such intensity variations are fairly narrow 
band $-$ decorrelation bandwidths $ \sim $ 100 kHz to a few MHz $-$ and exhibit 
modulations as large as 100\%.
This phenomenon, which has become known as Diffractive Interstellar Scintillation 
(DISS), has been extensively studied since the early days of pulsar observations
and is quite well understood (see Rickett (1977) for a review).  
DISS studies have been used  for probing the structure 
of electron density inhomogeneities in the ISM (e.g. Cordes, Weisberg \& Boriakoff 
1985; Armstrong, Rickett \& Spangler 1995) and to estimate pulsar velocities 
(e.g. Cordes 1986).

The discovery of long time scale ($ \sim $ days to months) flux variations 
(e.g. Cole, Hesse \& Page 1970; Huguenin, Taylor \& Helfand 1973)
and the subsequent correlation of these time scales with dispersion measure (DM)
(Sieber 1982) led to the recognition of a second class of propagation effects (Rickett, 
Coles \& Bourgois 1984), which has become known as Refractive Interstellar 
Scintillation (RISS).
In RISS, flux variations arise due to focusing and defocusing of the scattered
radiation by electron density irregularities that are large compared to the
Fresnel scale. These modulations are fairly broadband in nature.  Refraction through 
the large-scale density structures also produces the systematic ``drifting patterns''
that are often seen in pulsar
dynamic spectra (e.g. Smith \& Wright 1985; Hewish 1980).  RISS is also thought to
be responsible for the occasional occurrences of quasi-periodic intensity modulations
in the dynamic spectra (e.g. Wolszczan \& Cordes 1987; Hewish, Wolszczan \& Graham 1985).  
RISS is also the preferred explanation for other observed phenomena like slow flux variability 
at metre wavelengths (e.g. Rickett 1986), centi-metre 
wavelength ``flickering'' and discrete propagation events 
(e.g. Fiedler et al 1987) see with compact extra-galactic radio (EGR) sources
(see Rickett (1990) for a review).
Besides flux modulations, RISS is also thought to produce slow modulations 
of decorrelation bandwidth and scintillation time scale.  
It has been suggested that the large irregularities responsible for RISS are part 
of the same spectrum of irregularities which give rise to DISS (Rickett et al. 1984),
and studies of both phenomena provide us information on the spectrum over a very large 
range of spatial scales (several decades).  
While it is generally considered to be of a power-law form over the spatial scales of interest for ISS, 
the exact form of the spectrum, especially the slope, cutoffs, and above all, the validity of a simple 
power-law description, still remain to be well understood 
(cf. Narayan 1988; Rickett 1990; Armstrong et al. 1995).

On the theoretical front, several researchers have addressed the problem of DISS
and RISS in terms of small-scale ($ \sim $ $ 10^6 $ to $ 10^8 $ m) and large-scale 
($ \sim $ $ 10^{10} $ to $ 10^{12} $ m) electron density fluctuations producing two distinct 
regimes of scintillations 
(Rickett et al. 1984; Cordes, Pidwerbetsky \& Lovelace 1986; Romani, Narayan \& Blandford 1986).
Besides predicting the long-term flux variations
that characterize RISS, these models make predictions about the nature and levels of
fluctuations for observables such as decorrelation bandwidth, scintillation time scale and the drift 
rates of intensity patterns in pulsar dynamic spectra.  In the case of power-law 
models for the density irregularities, the refractive effects are expected to depend on the 
slope of the spectrum.  For example, the magnitude of the fluctuations of all the above 
parameters is predicted to increase substantially with the slope 
(Romani et al. 1986, Blandford \& Narayan 1985).

On the observational front, significant work has been done to measure the long-term flux 
modulations of several pulsars 
(Stinebring \& Condon 1990; Kaspi \& Stinebring 1992; Gupta, Rickett \& Coles 1993; LaBrecque,
Rankin \& Cordes 1994).
The results show that, for several pulsars, the measured
flux modulation indices are larger than those predicted by the simple Kolmogorov model, 
indicating that the underlying density spectrum may be more complicated.  Not much is known 
about the time scales and the magnitude  of the flux variation  due to RISS  and of other scintillation 
variables like decorrelation bandwidth and scintillation time scale.  
A recent study of the long-term variations in pulsar
dynamic spectra (Gupta, Rickett \& Lyne 1994) has shown that the properties of scintillation
patterns of several pulsars vary considerably with time, and to a first order,
these variations were found to be consistent with expectations from RISS.  
However, several discrepancies with the predictions of a Kolmogorov model were 
also observed in these data.  
Observations also show occurrences of unusual scattering effects such as episodes 
of multiple imaging and extreme scattering events (ESE), which are thought to be 
due to refraction by discrete structures in the ISM (e.g. Cordes \& Wolszczan 1986; 
Fiedler et al. 1987).
Most ESEs have been with EGR sources (Fiedler et al. 1994), and the only pulsar 
that is reported to have shown ESEs (in the form of unusual flux variations and timing
perturbations) is PSR B1937+21 (Cognard et al. 1993; Lestrade, Rickett \& Cognard 1998).
More recently, there have been a couple of attempts to test the quantitative predictions 
of theories and for two pulsars (PSRs B0329+54, B1937+214), the correlations between the 
fluctuations of scintillation observables were found to be consistent with the predictions
(Lestrade, Cognard \& Biraud 1995; Stinebring, Faison \& McKinnon 1996). 
However, the results are not conclusive and there is a need for more observations of this kind,
which is one of the motivations for the observations described in this paper.

The fluctuations of the flux and the DISS parameters caused by refractive scintillation lead to unreliable 
estimates of flux densities, scattering properties and  pulsar velocities, if obtained from only a 
few epochs of observations.
In order to get reliable estimates of these  quantities, the refractive fluctuations need to 
be averaged out by taking measurements at a large number of epochs spanning several refractive 
time scales.
The time scales of the fluctuations due to RISS, largely determined by the strength of scattering 
and the pulsar velocity, can range from a few days (for nearby pulsars at metre wavelengths) to 
several years (in the case of distant pulsars at longer wavelengths).
Most scintillation measurements made earlier 
(Roberts \& Ables 1982; Cordes et al. 1985; Smith \& Wright 1985; Cordes 1986)
could not take into consideration such effects (probably due to limited observing time),
and the scattering properties derived from such measurements are prone to errors due to 
refractive scintillation.
For example, in the published literature, estimates of decorrelation bandwidth and time 
by different observers differ by factors of three to five. 
A second motivation of the present work is to examine whether such discrepancies 
are attributable to RISS and to make a better determination of the scintillation parameters.

In order to investigate refractive scintillation effects and to obtain reliable estimates of 
scintillation properties of nearby pulsars, we undertook long-term systematic scintillation 
observations of eighteen pulsars using the Ooty Radio Telescope during 1993$-$1995.  
Our observations and measurements of the scintillation properties are presented in this paper 
(referred to as Paper I).  
In Paper II (Refractive effects and the spectrum of plasma density fluctuations), 
we use our results to constrain the electron density spectrum in the local ISM.
In Paper III (Testing theoretical models of refractive scintillation), we compare our results with 
existing theoretical models of RISS.
In an earlier paper (Bhat, Gupta \& Rao 1998), we have used our improved estimates of the average 
scintillation properties of several local pulsars to show that the distribution of scattering material 
in the local ISM is not homogeneous and that it supports the presence of the Local Bubble.

The layout for the rest of this paper is as follows.  Our observations are described in Section 2.
A description of the data analysis methods is given in Section 3, where we also present 
results on the diffractive and refractive scintillation properties.  We discuss the reliability 
of the present measurements in Section 4, where we also compare the present measurements with 
those from earlier observations and address the issue of long-term stability of scintillation 
properties and pulsar fluxes.
Section 5 gives a summary of our main conclusions from this work.


\section{Observations}

\subsection{Instrument and Data Acquisition}

The observations were made using the Ooty Radio Telescope (ORT), which is an equatorially mounted 
530 m x 30 m parabolic cylinder operating at 327 MHz (Swarup et al. 1971).
The telescope has an effective collecting area of 8000 $ m^2 $, system temperature of $ 150^o $ K and 
is sensitive to linearly polarized radiation with electric field in the North-South plane.
It has $ 9 { 1 \over 2 } $ hours of hour angle coverage and a declination coverage from $ - 55^o $ to 
$ + 55^o $. 
The ORT is a phased array, with 1056 dipoles at its feed, which was upgraded recently giving 
considerable improvement in sensitivity and stability (Selvanayagam et al. 1993).
With the current sensitivity, a signal-to-noise ratio of 25 can be achieved for a 1 Jy source,
for 1 sec integration and over a bandwidth of 4 MHz.
The signals from the dipole array are combined to form two signals from the North and South halves of 
the telescope.
These are input to a 1-bit correlation spectrometer (Subramanian 1989) to yield the cross power spectrum
of the signals from the two halves. 
For our experiment, pulsar data were taken over a bandwidth of 9 MHz centered at 327 MHz.
The data were obtained such that there are 64 frequency channels in the cross spectrum spanning
the observing bandwidth, yielding a frequency resolution $ \approx $ 140 kHz.

Pulsar data were acquired with a sampling interval of 6 msecs. 
The data were recorded both on the pulse and also on part of the off pulse regions.
The gated on pulse and off pulse 
regions were synchronously averaged in the correlation domain over a specified number of 
pulse periods and then recorded for off-line analysis.
A continuum source at a declination close to that of the pulsar being studied was observed for calibration.
The data from the calibrator were acquired for typical durations of 5 minutes and an equal stretch of data
were acquired by pointing the telescope to a cold region of the sky.
The calibration observations were generally made before starting the pulsar observations.
A suitable calibrator was chosen for each pulsar so as to eliminate any possible bias in the flux 
calibration.
The data from several calibrators taken on a given observing day were also used to monitor the stability
of the telescope gain over the observing periods.

\subsection{Sample Selection}

Our sample selection was made with the two-fold aim of 
(i) studying diffractive and refractive scintillations, and 
(ii) obtaining reliable estimates of scattering properties of nearby (distance $ \la $ 1 kpc) pulsars. 
It was largely determined by our preliminary calculations of expected scintillation parameters based 
on the earlier 
scintillation measurements available from the literature and the instrumental constraints.
For the pulsars that were known at the time of our observations, with 
DM $ < $ 40 \dmu and within 
the sky coverage of ORT, the expected values of decorrelation widths in frequency ($ \Delta \nu _{iss} $)
and in time ($ \tau _{iss} $) were calculated
from published results, assuming the scaling laws for a Kolmogorov density spectrum
($ \Delta \nu _{iss} ~ \propto ~ {\rm f_{obs} ^{4.4} } $, $ \tau _{iss} ~ \propto ~ {\rm f_{obs} ^{1.2} } $).
Pulsars whose scintillation patterns could be studied with a frequency 
resolution of 140 kHz over a 9 MHz band were selected.
The temporal resolution required to resolve the scintillation patterns in time is typically  10 secs
at 327 MHz.
In order to have sufficient signal to noise with this integration time, we included in our sample
only those pulsars whose flux at 327 MHz was greater than 25 mJy ($ S_{400} \la 20 $ mJy).
The minimum integration time with the correlation spectrometer being 6 msecs, we 
eliminated from our sample short period pulsars (period $ \la $ 100 msecs).
We also left out pulsars that are known to show intrinsic intensity variations (such as nulling)
over time scales comparable to that of ISS.
Pulsars for which no prior scintillation measurements were available, but satisfied our selection
criteria of DM, flux and period, were retained in the sample.
The final sample consisted of 18 pulsars in a DM range 5 to 35 \dmu, which are listed in Table 1 
along with relevant observational details.
Pulsar names are listed in column (2) and their DMs in column (3). 
Distance estimates adopted in our calculations (column (4))
are based on the electron density distribution given by Taylor \& Cordes (1993),
except for PSR B0823+26, for which we use the independent distance estimate available from 
parallax measurements (Gwinn et al. 1986).
The period of observation and the total number of epochs of observation are given in columns (5) and (6). 
The instrumental resolutions in frequency and time used for each pulsar are listed in columns (7) and (8).
The pulsar sample spans the DM range fairly uniformly and therefore allows us to study 
variations of diffractive and refractive properties over a range of DM, distance and strength of
scattering.

\subsection{Observing Strategy}

Each pulsar was observed at several epochs spanning many months.
For accurate estimates of diffractive scintillation parameters at each epoch,
observations need to be made over a time duration much larger than the
characteristic time scales ($ t_{obs} \gg \tau _{iss} $) and over bandwidths much larger than
characteristic frequency widths ($ B_{obs} \gg \Delta \nu _{iss} $) of patterns.
Since we had a fixed observing bandwidth of 9 MHz, 
pulsars which were expected to have large decorrelation bandwidths
($ \Delta \nu _{iss} $ $ \sim $ MHz) or decorrelation  times ($ \tau _{iss} $ $ \sim $ 1000 secs)
were observed for longer durations (typically $ \sim $ 2$-$3 hours) in order to ensure 
sufficiently large number of {\it scintles} needed to obtain good ensemble averages of 
scintillation properties. 
Pulsars with relatively shorter decorrelation widths either in time or in frequency,
were observed for shorter durations of about  $ 1-2 $ hours.
For refractive scintillation studies, dynamic spectra need to be monitored regularly for time 
spans much longer than their typical refractive time scales, with several
observations within the time scale. 
We estimated the expected values of refractive time scales
($ \tau _{r,exp} $) using the diffractive time scales available in the literature
in order  to decide the initial observing strategies for individual pulsars. 
The values of $ \tau _{r,exp} $ are given in column (9) of Table 1.
Since these time scales range from days to weeks,
observations were made over time spans of about $ \sim $ 100 days to ensure a sufficiently large number of
refractive cycles of fluctuations.
The final strategies in terms of number of epochs and their separations were largely 
influenced by the results from our own early observations.

Our final data are from four observing sessions during 
January 1993 to August 1995, with each session lasting over a period of 100 to 150 days. 
In every observing session, 6 to 8 pulsars were regularly monitored for their dynamic spectra.
The number of epochs of observations (\nep) has a wide range from 9 to 93, the reasons for which 
are as follows. 
Ten pulsars, with \nep $ \ga $ 20, were primarily observed for studying the refractive effects in 
dynamic spectra, while for the remaining nine, our basic aim was to obtain reliable estimates of 
scintillation properties for the purpose of studying the Local Interstellar Medium.
Four pulsars were re-observed in multiple sessions due to interesting scintillation properties 
observed during the initial sessions.
In Table 2, column (3) gives the number of observing sessions (\nsess) for each pulsar.
In columns (4)$-$(7), the break up of number of epochs for each session is given, where the quantity
shown in brackets is the time span of observation in days.
The total time span of observation (\tspmax) is given in column (8).
Pulsars PSR B0823+26 and PSR B1919+21 were observed over two sessions, the former due to unusual flux
variations that were seen and the latter due to the presence of persistent drift slopes. 
The basic interest in the case of PSR B1133+16, which was observed for three sessions, was the dramatic 
changes observed in the characteristics of its scintillation patterns.
Pulsar PSR B0834+06 was observed for the largest time span of observation (four sessions) to study 
persistent drift slopes seen in its dynamic spectra.
The data of these four pulsars also proved to be useful in investigating the long-term stability of 
both diffractive and refractive scintillation properties over time scales much longer than their
refractive time scales.


\section{Data Analysis and Results}

\subsection{Data Reduction Procedure}

\subsubsection{The Dynamic Scintillation Spectra}

To calibrate the pulsar data, data were acquired both on a nearby calibration source and 
on a cold region of the sky. This data were used to estimate the gain of the telescope
at the declination of the pulsar and to determine the bandpass of the correlation
receiver. For most observations, frequency channels beyond the 3 dB range ( about 7 MHz)
were not used. The off-source data were examined for bad channels and line interference
which were flagged for the pulsar data.

Pulsar data from the correlation spectrometer were edited for occurrences of 
instrumental malfunctions.
The data were calibrated for the telescope gain as well as gain variations across the
observing band.
The data were de-dispersed, examined for any systematic pulse drifts due to instrumental effects 
which were corrected if detected.
An integrated profile was obtained by averaging the pulsar data over the entire observing duration
and over the usable range of the observing bandwidth.
The on-pulse region was identified using a 5-$\sigma $ (where $\sigma $ is the rms of the off pulse region) 
threshold criterion above the mean off-pulse level.
The mean off-level was subtracted from the corresponding on-pulse region and the 
dynamic spectrum was obtained by integrating this over the on-pulse time bins.
A similar spectrum was created for the average value of the off-pulse region, which was used to detect 
regions of dynamic spectra corrupted by external interference.
The data were checked for both narrow-band and broad-band spurious signals, lasting over very short 
durations to the entire observing time, and the corrupted data regions were given zero weights in the 
further analysis.
The fraction of the data rejected in this manner seldom exceeds a few percent.
The pulsar flux density (F), averaged over the bandwidth and duration of the observations and the intensity 
modulation index ($m_d$) were computed from the on-pulse dynamic spectra.

\subsubsection{The Computation of Auto-covariance Functions}

In order to quantify the average characteristics of scintillation patterns at any epoch, 
we make use of the two-dimensional auto co-variance function (2-D ACF), which 
was computed for frequency lags up to half the observing bandwidth and for 
time lag up to half the observing time.
The function was corrected for the effect of receiver noise fluctuations and the residual, intrinsic 
pulse-to-pulse fluctuations, which remain uncorrelated and appear as a `ridge-like' feature at zero time
lag in the 2-D ACF.
We computed a weight function for the 2-D ACF which 
represents the uncertainties in the ACF values and is given by

\begin{equation}
\omega _d ( \nu , \tau ) ~ = ~ 
\sigma _n ~ \left( { N ( \nu = 0 , ~ \tau = 0 ) \over N ( \nu , ~ \tau ) } \right) ^{0.5}
\end{equation}

\noindent
where $ \nu $ and $ \tau $ are the frequency and time lags.
$ N ( \nu , \tau ) $ is the number of data pairs averaged in computing the ACF value at $ ( \nu , \tau ) $.
The quantity $ \sigma _n $ represents the rms noise in the ACF at $ \nu = 0 $ and $ \tau = 0 $,
and was computed from the region of the ACF where interstellar features are absent.
This weight function was used while fitting the 2-D ACF with  suitable functions.


\subsection{Description of Data}

Sample dynamic spectra are shown in Figs. 1(a)$-$1(h) and in Figs. 3(a)$-$3(m).
These spectra have been selected to illustrate the general characteristics of pulsar scintillation
as well as to highlight the observed variations from pulsar to pulsar and also from epoch to epoch.
In Figs. 1(a)$-$1(h), there are a number of panels, each displaying the spectrum 
at a given epoch.
Multiple panels are shown mainly to highlight the changing form of dynamic spectra with time.
The observing durations are typically  2$-$3 hours, 
but often the displays have been restricted to shorter
durations ($ \sim $ 100 minutes), as this is sufficient to illustrate the basic features.
Most data have resolutions of 140 kHz in frequency and  10 secs in time.
The date of observation is indicated at the top right corner of each panel and
the mean flux density for each epoch is shown at the top left corner.
In the gray-scale representation of the intensity, darker regions correspond to higher intensity values
and lighter regions to lower values.
The display saturates to black at four times the mean intensity and white regions are usually at about 20\% 
of the mean.
There are bright intensity regions, usually known as {\it scintles}, which are resolvable when their 
widths in frequency and time are larger than the instrumental resolution.
A typical spectrum consists of a large number of such {\it scintles} of varying intensity, 
shape and widths.
Such random, deep modulations of intensities, occurring over narrow frequency ranges and short time 
intervals are general features of all spectra. 
Various time dependent  instrumental problems as well as external broad band interference that have been 
identified and blanked out in our data reduction process appear as vertical white regions that are 
distinguishable from regions of real flux fading by their sharp boundaries.
The horizontal white strips seen in some of the spectra 
(for example in Figs. 1.b and 1.g, ie., pulsars \egta and \ninea) are regions of the data corrupted by 
narrow band interferences.
However, as seen in the figures, only a very small fraction of the data is corrupted 
by such spurious signals and therefore this does not affect the estimation of 
the scintillation parameters.

Sample ACFs of some selected pulsars are displayed in Figs. 2(a)$-$2(c), to illustrate the
general characteristics of ACFs and to highlight the special features seen with some of the pulsars.
Each figure has a  number of panels representing the ACF of the dynamic spectrum obtained
on the observing day indicated at the top of the panel.
The displays are restricted to much smaller ranges in frequency and time lags than the maximum
lag ranges for which the ACFs have been computed.

The sample data shown here (Figs. 1, 2 and 3) illustrate the diversity seen in the pulsar dynamic 
spectra.
The properties of patterns, such as their sizes and shapes in the time-frequency plane, vary 
significantly from pulsar to pulsar. 
The widths of intensity patterns vary from $ \sim $ 100 secs (e.g. PSR B0823+26, PSR B0919+06, PSR B2045$-$16)
to as large as $ \sim $ 1000 secs (e.g. PSR B1604$-$00, PSR B2016+28) in time and from $ \sim $ 100 kHz 
(e.g. PSR B0329+54, PSR B1540$-$06, PSR B2310+42) to several MHz (e.g. PSR B1133+16, PSR B1237+25,
PSR B1929+10) in frequency.
Organized drifting of scintles in the frequency-time plane is seen in the dynamic spectra of several pulsars.
The data of PSR B0834+06, PSR B1133+16, PSR B1919+21 and PSR B2045$-$16 are some of the best 
examples with prominent drifting patterns.

The changing form of dynamic spectra with epoch is a common feature for all pulsars and significant 
variations are seen over time scales as short as a few days.
The property is better illustrated  through the plots of ACFs, whose widths and orientations 
show significant variations from epoch to epoch.
Significant variations are also present in the average flux densities of pulsars.
Generally, at a given epoch of observation, most of the intensity patterns appear with quite similar 
orientations in the time-frequency plane.
But there are slow variations in the magnitudes of tilts from epoch to epoch, which is seen as a 
changing nature of elongations of the ACFs. 
For many pulsars, pattern tilts undergo sign reversals over time scales of the order of several days.
Pulsars PSR B0823+26 and PSR B0919+06 are good examples showing such systematic variations of pattern 
slopes along with sign reversals.

Systematic slope variations and sign reversals are found to be common for most of the pulsars observed.
There are, however, some pulsars whose data are characterized by a very few or 
even an absence of sign reversals of pattern slopes over the observing time spans 
(typically $ \sim $ 100 days).
Pulsars PSR B0834+06 and PSR B1919+21 are best examples of such ``persistent drift slopes''. 
PSR B1919+21 is a unique case, where in addition to the usual intensity decorrelations, 
there are deep intensity modulations lasting over very short time scales (roughly one minute).
This broadband phenomenon is found to be a stable feature of this pulsar over the entire observing
session.
Another special pulsar is PSR B1133+16, which shows remarkable changes in the nature of the 
dynamic spectrum between its successive observing sessions.
On 30 April 1994 this pulsar shows periodic intensity modulations in the dynamic spectrum.

In the rest of this section, we briefly describe the observed properties of some of the pulsars
that show typical behaviour and some with anomalous scintillation properties.
This is followed by a brief summary of the general characteristics of the remaining pulsars.


\subsubsection{PSR B0823+26}

Sample data of this pulsar are shown in Fig. 1.a 
which has 9 dynamic spectra spanning a period of 57 days from the second observing session 
(October 1993$-$January 1994), and in Fig. 2.a which  shows 9 selected ACFs spanning 54 days from the 
first observing session (March 1993$-$May 1993).
These data illustrate a number of basic observable effects due to diffractive and refractive scintillations.
The intensity scintillation patterns arising from rapid intensity modulations in time and frequency are
clearly seen.
The patterns decorrelate over a few 100 kHz in frequency and $ \sim $ 100 secs in time for this pulsar.
Often, the patterns appear with preferred orientations in the time-frequency plane and, on a given day,
most of them exhibit slopes of same sign $-$ either positive or negative.
The sloping features in the dynamic spectra produce tilted contours in the ACFs.
In addition, there are fine variations of sizes and shapes of patterns from day to day,
along with significant variations in the average flux density.
The property is better illustrated in Fig. 2.a, where ACFs at various epochs differ in terms of 
their sizes and elongations.
A systematic slope reversal can be seen in the 6 ACFs starting from 26 April 1993.

\subsubsection{PSR B0834+06}

Fig. 1.b show the dynamic spectra for this pulsar, 
at 25 epochs spanning a period of 110 days, taken from the first observing session (January 1993$-$May 1993).
A similar data set (at 23 epochs spanning 100 days) 
from the final observing session (April 1995$-$July 1995) is displayed in Fig. 1.c.
These data have been selected to illustrate a number of special properties seen with this pulsar.
Drifting patterns are highly pronounced in the dynamic spectra of this pulsar (Fig. 1.b).
Scintles are much broader than that of PSR B0823+26, 
with typical widths of $ \sim $ 500 kHz in frequency and $ \sim $ 400 secs in time.
The variable nature of dynamic spectra is also much better pronounced here.
In the data taken from the first session (Fig. 1.b),
patterns appear with slopes of same sign for all the epochs and do not show 
any reversal in between.
Such persistent drifting features are also seen in the data from the next two observing sessions
$-$ October 1993 to January 1994 and February 1994 to June 1994.
However, quite dramatic changes are noticed in the data taken during the final session $-$
April 1995 to July 1995 (Fig. 1.c), in which there are several occasions of slope reversals.
In addition, there are a few examples of spectra with `dual slopes' (e.g. 10 June 1995).

\subsubsection{PSR B0919+06}

Selected dynamic spectra and ACFs of this pulsar are shown in Figs. 1.d and 2.b where
the data at 9 epochs spanning a period of 35 days are displayed.
The behaviour of this pulsar is quite similar to that of PSR B0823+26.
Intensity patterns fade over a few 100 kHz in frequency and $ \sim $ 100 secs in time
and the widths  vary significantly between successive epochs of observation.
Gradual and systematic variations in the slopes of the patterns 
are highly pronounced here.
In fact, the orientation of the major angle of the ACF reverses its sign twice $-$ from 28 April 1994 to 14 May 1994 and then 
again from 16 May 1994 to 1 June 1994.
In some ACFs there is evidence  for a dual slope where the orientation is different for the higher and 
lower contours - data on 4 May, 14 May
and 24 May are good examples.
Such behaviours can be expected during the transition periods of drift reversals,
which is clearly substantiated by the dynamic spectra.
This means, for this pulsar, pattern slopes seem to undergo a sign reversal typically once in 20 days.

\subsubsection{PSR B1133+16}

Another special pulsar is PSR B1133+16, which shows remarkable changes in 
properties of its dynamic spectra from one observing session to another.
The pulsar was observed for 3 well separated sessions over a 3-year period.
Fig. 1.e displays the dynamic spectra of this pulsar from the observations 
during March 1994$-$June 1994.
A similar data set from the third session (April 1995$-$July 1995) is shown in Fig. 1.f.
Each figure consists of data from 16 epochs of observations.
Data shown in Figs. 1.e and 1.f span 65 and 92 days respectively.
A typical dynamic spectrum from the initial observing session (February 1993) is shown in Fig. 3.c, which is
characterized by intensity patterns fading over $ \sim $ a few 100 kHz in frequency and over $ \sim $ 
100 secs in time.
In contrast, dynamic spectra taken during the second session (Fig. 1.e), in particular those from the first
half of the session, show patterns that fade over much broader ranges in frequency ($ \sim $ 1 MHz) 
and in time ($ \sim $ 200 secs).
The spectrum on 30 April 1994 shows clear evidence of periodic intensity
modulations in the frequency-time plane, known as `interstellar fringes', which result from
multiple imaging caused by refraction in the ISM (e.g. Wolszczan \& Cordes 1987).
It may be mentioned that such imaging events were reported earlier for this pulsar 
by Hewish et al. (1985) and Cordes \& Wolszczan (1986).
After this episode of fringes, the pulsar appears to return to the initial mode 
seen in session I, with intensity patterns fading over $ \sim $ a few 100 kHz in frequency and over 
$ \sim $ 100 secs in time.
This mode of scintillation spectrum prevails until the last observing day of the session
(1 June 1994).
Surprisingly, the dynamic spectra data obtained from the subsequent observing session (Fig. 1.f) 
are markedly different from the earlier data,
showing intensity patterns fading over much broader ranges in frequency 
($ \sim $ a few MHz), which is similar to the characteristics of the first half of session II, 
and sometimes seem to be broader than the observing bandwidth itself.
The pulsar thus seems to exhibit drastic changes in the nature of its dynamic spectrum 
on time scales $ \sim $ 1 year.
Behaviour of this kind is not seen for any other pulsar that was followed-up for multiple
observing sessions.

\subsubsection{PSR B1919+21}

Selected dynamic spectra  of PSR B1919+21 from the first observing session (March to May 1993) 
are shown in Fig. 1.g, while in Fig. 2.c are shown selected ACFs from the follow-up session 
(October 1993 to January 1994).
The spectra in Fig. 1.g are from 9 observations spanning 23 days while the
ACFs displayed in Fig 2.c are for 25 observations over a period of 87 days.
These data have been selected to highlight some of the special properties seen with this pulsar.
The dynamic spectra appear to be quite similar to that of PSR B0834+06 shown in Fig. 1.b,
but substantial variations are seen in size and shape of ACF over a time interval as short as 1$-$2 days.

A noticeable property seen in the dynamic spectra of this pulsar is fine, deep modulations 
of intensity patterns, which are invariably broadband and last over $ \sim $ 100 secs in time.
The property is unique to this pulsar and is a stable feature in the data from both the 
observing sessions.
These intensity modulations produce a narrow ridge in the ACF at zero time lag (see Fig. 2.c).
Sometimes, this effect also leads to fine corrugations of the ACF along the  time axis
(see for example ACFs on 22 October 1993 and 29 November 1993).
Because of its broadband nature and persistence over entire duration of observation, we believe that these
modulations are intrinsic to the pulsar and are not a scintillation phenomenon.
Thus, the ACFs of this pulsar are  more complex than those of the rest of the pulsars.

Drifting patterns are highly pronounced in the dynamic spectra of this pulsar (Fig. 1.g) as is 
the persistent nature of drift slopes.
The data from the follow-up observations (Fig. 2.c) during October 1993$-$January 1994 also show 
persistent tilts in the ACF.
There seems to be occasional instances of reversals of drift slopes, but
the overall appearance of the data are more like that of persistent drifts.
No further observations were made and hence no information is available on how long the persistent 
drifting features last for this pulsar.

\subsubsection{PSR B2045$-$16}

Sample dynamic spectra of this pulsar are displayed in Fig. 1.h, 
which consists of observations at 9 epochs spanning a period of 40 days.
The pattern drifts are quite prominent as in PSR B0834+06 and PSR B1133+16, 
and show large variations and a number of reversals of slope over a period of 3 months.
In contrast with the data discussed so far, the intensity patterns of this pulsar 
show random occurrences of broad band intensity fluctuations lasting for a few time samples 
($ \sim $ 10 secs), thereby giving rise to white vertical strips in its dynamic spectra.
Broadband features of this kind are due to an insufficient averaging of pulsar's intrinsic 
intensity variations due to the presence of additional phenomena such as nulling.


\subsubsection{Summary of the Remaining Pulsars}

The data of PSR B2020+28 (typical spectrum is shown in Fig. 3.k)
show a behaviour quite similar to that of PSR B0919+06 and PSR B0823+26.
The data of pulsars PSR B0628$-$28 and PSR B2327$-$20 are also found to be quite similar to 
that of PSR B0823+26 and PSR B0919+06, 
except that patterns fade over much longer durations, typically $ \sim $ 500 secs.
Typical spectra of these pulsars are shown in Fig. 3.b and Fig. 3.m respectively.
There are frequent, broad band striations in the spectrum of PSR B2327$-$20, 
which are presumably due to residual intrinsic intensity variations.
Typical spectra of PSR B1604$-$00 and PSR B2016+28 are shown in Figs. 3.g and 3.j, where
the intensity patterns fade over much longer durations in time, typically $ \sim $ 1000 secs.
The sloping patterns are not quite noticeable in the spectra of PSR B2016+28.
In contrast, PSR B1604$-$00 shows prominent drift slopes, which vary systematically and show a sign
reversal in between.
The spectra of this pulsar are rather poorly sampled and there are only 10 epochs of observations
over a time span of 94 days.
Fig. 3.d and 3.i show typical spectra of pulsars PSR B1237+25 and PSR B1929+10.
For these pulsars, patterns are much broader in frequency, and fade over a range $ \sim $ 1 MHz.
Often, their spectra are characterized by a few bright scintles that dominate over others and 
last for $ \sim $ 1000 secs, which are followed by long fading over similar time scales.
The presence of bright scintles also cause apparent intensity modulations that are somewhat
greater than 100\%.
In the case of pulsars PSR B0329+54, PSR B1540$-$06 and PSR B2310+42, typical spectra of which are shown 
in Figs. 3.a, 3.f and 3.l respectively, the intensity patterns fade over a very narrow frequency range, 
sometimes as small as $ \sim $ 100 kHz, which is close to our spectral resolution limit.
The patterns are thus significantly smoothed in frequency due to the instrumental resolution and this 
smoothing effect reduces apparent intensity modulations to somewhat below 100\%.
The spectra of pulsars PSR B1508+55, PSR B1747$-$46 and PSR B2310+42 (Figs. 3.e, 3.h and 3.l respectively)
are characterized by poor signal-to-noise ratio because
of a significant reduction in the telescope gain at high declinations.


\subsection{Estimation of Scintillation Properties}

\subsubsection{ACF Fitting and Estimation of Scintillation Parameters}

The 2-D ACF can be characterized by its widths along the frequency lag and time lag axes and its
orientation in the frequency lag-time lag plane.
The parameters corresponding to these widths are the decorrelation bandwidth, \nd, defined as 
the half-power width along the frequency lag axis at $\tau = 0$, and the scintillation time, \td, 
defined as the $ {\rm e ^{-1} } $ width along the time lag axis at $\nu = 0$.
The 2-D ACF is fitted with a two-dimensional elliptical Gaussian function (see Gupta et al. (1994) 
for justification) of the following form.

\begin{equation}
\rho _{_k} ^m (\nu , \tau ) ~ = ~ C_o ~ exp \left[ - \left( C_1 ~ \nu ^2 ~ + ~ C_2 ~ \nu ~ \tau ~
+ ~ C_3 ~ \tau ^2 \right) \right]
\end{equation}

The amplitude of the Gaussian function, $ C_o $, is unity, since the ACF itself is normalized to unity.
In the fitting algorithm, the deviations between the ACF and the model Gaussian are weighted by 
their uncertainties given by the weight function, $ \omega _d $, as described in equation (1).
This weighting scheme accounts for the measurement uncertainties in the ACF values due to finite 
data stretch, and does not take care of the `scintle-lumpiness' effect arising due to finite number 
of scintles (this is addressed later in this section).
The model Gaussian parameters, $ C_1 $, $ C_2 $ and $ C_3 $, are estimated by a $ \chi ^2 $-minimization 
procedure.
The scintillation parameters \nd and \td are obtained from these fitted parameters as 

\begin{equation}
\nd ~ = ~ \left( { ln ~ 2 \over C_1 } \right) ^{0.5}
\hspace{1.0cm} ; \hspace{1.0cm}
\td ~ = ~ \left( { 1 \over C_3 } \right) ^{0.5}
\end{equation}

The decorrelation widths \nd and \td obtained in this manner are corrected for smearing due to finite
instrumental resolutions in frequency and in time using a quadrature subtraction scheme.
Barring a few exceptions, such as \nd measurements of pulsars PSR B1540$-$06 and PSR B2310+42,
this correction is not significant for our measurements.

To characterize the drifting features in the dynamic spectra 
(or equivalently, the tilt or orientation of the fitted Gaussian $ \rho _k ^m $),
two issues need to be addressed.
The first is the choice of an appropriate quantity for describing the effect.
Usually the sloping features have been characterized by the frequency drift rate, measured as \dnt
(Hewish 1980; Smith \& Wright 1985).
More recently, the inverse of this quantity, ie., \dtn, has been suggested as a more appropriate 
choice (Spangler 1988; Rickett 1990; Gupta et al. 1994), as it has a more meaningful connection with theory.
\dtn is proportional to the refractive scattering angle \refr, and will have the
same statistical properties $-$ zero mean random variable, in the simplest case.
We therefore prefer to use \dtn in place of \dnt in characterizing drift slopes.

The second issue concerns the correct method of estimating \dtn.
At first sight, it may appear that the slope of the major axis of the ellipse fitted to the ACF is a
good measure of \dtn.
However, this is not true since the ellipticity and hence the slope of the major axis depends 
on the plotted scale and can give results that are not meaningful 
when the major axis happens to be aligned along one of the axes.
The basic effect of refraction due to a linear phase gradient is not a tilting
of the entire pattern, but rather a `shear' resulting from the frequency-dependent displacements of patterns
in the observing plane.
Therefore, in the quantity \dtn, $ dt $ should refer to the time interval corresponding to the differential 
displacement of intensity patterns separated in frequency by $ d \nu $.
The proper measure of \dtn would be the slope of the line joining the points on the ellipse with the highest 
correlation at a given frequency offset.
This definition is similar to that suggested by Gupta et al. (1994).
It results in a zero drift slope in the absence of sloping patterns and is also free 
from the ambiguity of determining the major axis.
In terms of the fitted  Gaussian parameters, this slope can be expressed as

\begin{equation}
{ d t \over d \nu } ~ = ~ - \left( { C_2 \over 2 ~ C_3 } \right)
\end{equation}

The uncertainties in $ C_1 $, $ C_2 $ and $ C_3 $ due to the Gaussian model fitting 
are obtained from the $ \chi ^2 $ analysis
and translated into corresponding uncertainties in \nd, \td and \dtn,
and are referred to as \smod.
We also take into account the statistical uncertainties in the scintillation parameters arising due to
finite number of scintles, given by

\begin{equation}
\sigma _{est} ~ = ~ \left[ f_d \left( { B_{obs} ~ t_{obs} \over \nu _d ~ \tau _d } \right) \right] ^ {-0.5}
\end{equation}

\noindent
where $ \sigma _{est} $ is the fractional error and $f_d$ is the filling fraction for number of scintles, 
which is assumed to be 0.5 in our calculations.
This is just a moderate value, and may overestimate the number of scintles 
(hence underestimate \sest)
if typical separation between the scintles is much larger than their sizes.
For the parameters \nd, \td and \dtn, errors from the Gaussian fitting ($ \sigma _{mod} $)
are added in quadrature with the
statistical errors to get their final uncertainties.
The typical values of $ \sigma _{mod} $ and $ \sigma _{est} $ correspond to errors of 
10\% and 5\% respectively.

Similar statistical errors are also applicable for the measurements of pulsar flux densities and 
intensity modulation indices, which are directly obtained from the dynamic spectra.
The intensity modulation index, which is known as the diffractive scintillation index, $m_d$, 
is usually found to be close to unity, within the measurement uncertainties.
Moreover, the measured fluctuations of $m_d$ are found to be comparable to the uncertainties in $m_d$ values,
and hence $m_d$ can be treated as a stable quantity.
The observed 100\% intensity modulations are in accordance with
the strong scintillation expected for pulsars at metre wavelengths.
However, there are a few exceptions (such as PSRs B1237+25 and B1929+10) with significant deviations 
of indices from unity and they are found to result from dominance of a few bright scintles.
Occasionally, indices greater than  unity are seen when a few bright scintles dominate the dynamic spectrum.
However, no pulsar shows any systematic, large variations of $m_d$.


\subsubsection{Time Series and Average Values of Scintillation Parameters}

The results from the analysis are presented in the form of time series of four quantities:
the three scintillation parameters \nd, \td and \dtn, and average flux density (F).
These are shown in Figs. 4(a)$-$4(x), where results for each session of every pulsar are 
shown as a separate panel,
which is divided into 4 sub-panels showing variations of the four quantities.
In each figure, day number 1 corresponds to the starting day of observation.
When the pulsar was observed for more than one session, the results 
are shown separately since we find that the successive sessions show
significant difference in the scintillation properties.
Thus there are four panels for PSR B0834+06 and two each for PSR B0823+26, PSR B1133+16 and PSR B1919+21
(see Table 2 for details).
The uncertainties of individual measurements are estimated as described in the previous section.
The length of the error bar represents $\pm$1-$\sigma $ uncertainty in the measurement.
The data points are joined with dotted lines merely to illustrate the general trends seen in the 
variations of the quantities.
We now use these time series to derive the average properties characterizing the diffractive and 
refractive scintillations for each pulsar.

The parameters \nd and \td form two basic observables of diffractive scintillation that are measurable 
from our data.
The present observations show that  they vary significantly with time, which is presumably due to refractive 
effects.
We discuss the details of the modulation characteristics in a later section.
The important point here is that scintillation measurements from only a few observations would 
lead to erroneous conclusions about average scattering parameters.
Most earlier measurements in the literature could not take into consideration such effects, probably
due to limited observing time.
It is necessary to average out these fluctuations in order to get reliable estimates of the quantities.
Our observations were made over time spans much longer than the expected time scales of fluctuations and 
we were able to obtain many more scintillation measurements on each pulsar in our program.
Thus the data allow us to reduce the errors due to refractive scintillation effects, as a result of which
we are able to estimate DISS parameters more robustly than previous attempts.
The average values thus obtained $-$ \avnd  and \avtd $-$ are listed in columns (2) and (3) of Table 3.
It may be mentioned that our data span may be insufficient to yield stable ISS parameters for pulsars 
which show persistent drifting features lasting over many months. 

The third parameter \dtn is basically related to refractive scattering.
Nevertheless, we briefly describe it here since the quantity is measured from the dynamic spectra,
which is due to DISS.
Like \nd and \td, the drift slopes of intensity patterns also show significant variations with time.
The theoretical expectation is that these slopes should vary randomly about a zero mean over 
refractive time scales.
In general, most pulsars show this basic property where the slopes change their magnitudes and
show frequent sign reversals.
However, PSR B0834+06 and PSR B1919+21 seem to be exceptions; the former does not show sign reversals
in the first three sessions, and the latter shows quite similar property though there are a few epochs of 
opposite drift slopes.
Thus, excepting a few pulsars, the mean value of the drift rate \avdtn (given in the column (4) of Table 3) 
is zero within the estimation errors.

\subsubsection{The Global ACF Computation}

Here we describe an alternative method which gives more reliable estimates of average values of \nd, \td and 
\dtn than obtained by averaging the time series.
The method makes use of a weighted average 2-D ACF, which we refer to as the Global 2-D ACF (GACF).
It is computed from the ACFs at all the epochs of observation for a given pulsar using the definition

\begin{equation}
\rho _g \left( \nu , \tau \right) ~ = ~ { \sum _{k=1} ^{k=N^{'}_{ep}} ~ \varpi _k \left( \nu , \tau \right) ~
\rho _k \left( \nu , \tau \right) \over
{ \sum _{k=1} ^{k=N^{'}_{ep}} ~ \varpi _k \left( \nu , \tau \right) } }
\end{equation}

\noindent
where $ \rho _k $ is the ACF of dynamic spectrum at $ k^{th} $ epoch.
$ N^{'}_{ep} $ is the number of observations made with identical resolutions in time and frequency.
$ \varpi _k $ is the weight function for the GACF, which is simply the number of data pairs averaged 
to get $ \rho _k$ (note that it is different from the quantity, $ \omega _d $, as defined in eq. [1]). 
Estimating the average scintillation parameters from the GACF 
has the following advantages over the time series averages.
First, the individual ACFs are given weights proportional to their statistical reliabilities, which would
make the average values less sensitive to ACFs computed from very short data stretches. 
The GACF has higher signal-to-noise ratios than the individual ACFs  and is less sensitive to outliers, 
which results in smaller uncertainties in the fitting procedure.
Further, while computing the GACF, deviations of the individual ACFs from their model gaussians 
get averaged out and so the  GACF will be much closer to its model gaussian form than the
case of individual ACFs.
The uncertainties due to the model fitting (\smod), which are largely due to the deviation
from the gaussian shape, will be smaller than that in the case of normal ACFs. 
Thus, the estimates of average values obtained from this technique will be more robust than those from the
time series.
The only limitation is that the data should be taken with identical
resolutions in time and frequency, which is the case with most of our data.

The GACFs obtained from our data are shown in Fig. 5.
For pulsars with multiple sessions of observations, GACF is computed for each session separately.
In some cases, part of the data in a given session was obtained with a different resolution either 
in time or in frequency, in which case GACFs are computed for the corresponding data sets separately.
In such cases, the labels A and B are attached alongwith the session ID (e.g. PSR B0823+26(IA)) or 
pulsar name (e.g. PSR B1508+55(A)) to distinguish between the different parts of data.
The GACF is fitted with a gaussian of the form described in equation [2] to yield parameters \ndg, \tdg and 
\dtng $-$ which are the average estimates of \nd, \td and \dtn respectively.
These are presented in columns (5), (6) and (7) respectively of Table 3.
The values are, in general, comparable to the corresponding time series averages.
However, there are some exceptions.
For PSR B1237+25, the two methods give significantly different values.
The estimates of \ndg and \tdg are larger than the time series averages. 
But, interestingly, the drift slope obtained from the GACF is very close to zero,
in comparison to a significant value indicated by the time series.
A similar property is also seen for PSR B1929+10.
For \eleva, both methods yield significant average drift slopes, which may be due to its poor 
statistics (\nep = 6).
The discrepancy between the \nd values of PSR B2310+42, could be due 
its being close to the frequency resolution of our setup.
This can be clearly seen from 
the time series, where \nd values corrected for instrumental smearing become lower than the resolution
in frequency.
For PSR B2327$-$20, the discrepancy seen between \avtd and \tdg seem to be due to some `outlier' data points
in the initial epochs of observation.
We believe the results from the GACF technique 
to be more robust than those from the time series and hence use them in our further analysis in this paper.


\subsubsection{Estimates of Derived Scattering Parameters and Pulsar Flux Densities}

Using the results obtained from the GACF method, we estimate some quantities characterizing the strength of 
scattering and scintillation pattern speed.
The  conventional way of describing the strength of scattering is in terms of
\cn (e.g. Rickett 1977; Cordes et al. 1985), where 
the spectrum of the plasma density fluctuations is considered to be a power-law, given by 
$ {\rm  P _{\delta n_e} (\kappa) = \cn ~ \kappa ^{-\alpha}  } $, where $ \kappa $ is the spatial 
wavenumber.
The quantity \cn is a measure of rms electron density fluctuations which give rise to scattering. 
The line-of-sight average of \cn is given by (Cordes et al. 1985), 

\begin{equation}
{\rm \avcn ~ = ~ 2 \times 10^{-6} ~ \left( f _{obs,MHz} \right)^{11/3} ~ \left( D_{pc} \right)^{-11/6} ~
\left( \nu _{d,kHz} \right)^{-5/6}  
\hspace{1.0cm}
m^{-20/3}}
\end{equation}

\noindent
for a Kolmogorov form of density spectrum ($ \alpha = 11/3 $).
Our estimates of \avcn are given in column (3) of Table 4, where the uncertainties are due to 
the measurement errors in \ndg.
In all our calculations, we make use of the most recent pulsar distance estimates available from 
Taylor, Manchester \& Lyne (1993),
except for PSR B0823+26, for which an independent distance ($ \approx $ 380 pc) from the parallax method 
is available (Gwinn et al. 1986).
The new, improved  estimates of \avcn derived from our measurements have been used to study the distribution 
of scattering material in the local interstellar medium (Bhat et al. 1997, 1998).

Another quantitative measure of the strength of scattering is the 
parameter $u$
which  is defined as the ratio of the fresnel scale (\rf) to the `coherence scale' 
(\so = $ (k \theta _s)^{-1} $).
In terms of decorrelation bandwidth, this quantity can be expressed as (Rickett 1990; Gupta et al. 1994)

\begin{equation}
u ~ = ~ \left( { 2 ~ \fobs \over \nd } \right) ^{0.5}
\end{equation}

\noindent
The values of $u$ obtained from \ndg are given in column (4) of Table 4 and they range from 20 to 75. 
The observable effects of refractive scintillation are thought to critically  dependent on this parameter, and 
the condition $ u > 1 $ is considered to be the strong scattering regime.
The quantity \avcn is representative of average scattering along the line-of-sight, 
while $u$ is an indicator of the integrated scattering at a given frequency of observation.

The scintillation pattern speed, \viss, which measures the relative motion between the scintillation pattern 
and the observer, is estimated from measurements of decorrelation bandwidth and scintillation time scale
(e.g. Cordes 1986), and is given by 

\begin{equation}
\viss = ~ \Av \left[ { \sqrt { D_{kpc} ~ \nu _{d,MHz} } \over 
\left( f_{obs,GHz} ~ \tau _{d,secs} \right) } \right]
\hspace{1.0cm}
\velu
\end{equation}

\noindent
For the constant \Av, we adopt the value $ 3.85 \times 10^4 $ given by Gupta et al. (1994).
In the above expression, we have considered a thin screen placed at the midway between the pulsar 
and the observer.
Values of \viss derived from measurements of \ndg and \tdg are given in column (5) of Table 4, 
with uncertainties that reflect the measurement errors in \ndg and and \tdg.
For pulsars PSRs B1747$-$46, B2310+42 and B2327$-$20, scintillation speeds are reported for the first time.
Though \viss represents the net effect of pulsar's proper motion (\vprop), Earth's orbital motion around 
the Sun (\vobs) and bulk flow of the density irregularities in the ISM (\virr), contributions 
due to \vobs and \virr are usually small in comparison to that due to \vprop.
Therefore, scintillation speeds are generally considered to be useful indicators of pulsar's proper 
motion speeds.

The long-term nature of our observations have enabled us to obtain reliable flux density values by 
averaging out the fluctuations due to refractive scintillation.
The average flux densities (\Sort) are given in the column (7) of Table 4, and have measurement errors 
of the order of  5\%. 
The uncertainties are due to absolute flux calibration and estimation error due to finite number of 
scintles.
Since the ORT is sensitive to linearly polarized radiation with electric field in the local North-South plane, 
our flux densities may turn out to be underestimates, especially for pulsars which have significant 
linearly polarization.


\section{Discussion}

We have presented results from a long-term systematic study of the scintillation properties of a large
number of pulsars.
Before we compare these average scintillation measurements with earlier studies and discuss the long-term 
stability of the scintillation parameters,
we estimate the various errors in our computations of average scintillation properties.
Two main sources of errors are 
(i) errors due to finite number of independent epochs of observations, which we denote as \smeas, and 
(ii) errors due to limited number of refractive cycles of fluctuations spanned by the measurements, 
denoted as \sstat.
A direct estimation of the latter is not practical; hence, we get first order estimates by assuming a 
stationary statistics for RISS. 
This assumption may not be true for some pulsars, in which case our values of \sstat will be less reliable.
In Appendix A, we describe the details of our scheme for estimating these errors.
The results of our error calculations are summarized in Table 5, which gives the percentage errors for 
\ndg, \tdg and \avfd.
In general, the second kind of errors dominate over the first kind; typical values of \smeas are 
$ \sim $ 2$-$5\%, while that of \sstat $ \sim $ 5$-$15\%.
Overall, our data represent a significant improvement in terms of accuracy and reliability of scintillation 
measurements compared to those available in the literature, 
most of which were obtained from observations at a few epochs.
We now compare the present measurements of diffractive scintillation parameters (\ndg and \tdg)
and flux density (\avfd) with earlier measurements.
We also examine the agreement between our estimates of scintillation speeds and the proper motion 
speeds of pulsars.


\subsection{Comparison with Earlier Measurements}

\subsubsection{Decorrelation Bandwidth and Scintillation Time Scale}

Measurements of decorrelation bandwidth and scintillation time scale reported in five papers 
$-$ Cordes (1986), Cordes, Weisberg \& Boriakoff (1985), Smith \& Wright (1985), Roberts \& Ables (1982)
and Gupta et al. (1994) $-$ have been scaled to our observing frequency assuming scaling laws 
$ \nd ~ \propto ~ {\rm f_{obs} ^{4.4} } $ and $ \td ~ \propto ~ {\rm f_{obs} ^{1.2} } $, which correspond 
to a power-law spectral index $ \alpha $ = 11/3.
The \nd and \td estimates calculated in this fashion are given in Tables 6A and 6B.
While the measurements of Gupta et al. (1994) are from similar observations, ie., from several epochs over 
a long period, the remaining ones are from a fewer epochs. 

As seen from Table 6, there is a general lack of agreement between various measurements.
Since refractive scintillation is thought to be the cause of slow fluctuations of diffractive observables,
we first examine how well the discrepancies can be accounted for in terms of RISS-induced fluctuations.
For pulsars PSR B1929+10 and PSR B2045$-$16, the differences between the measurements are within the rms 
fluctuations seen in our time series for \nd and therefore can be explained in terms of RISS.
However, the discrepancies are much larger for the remaining pulsars.
If we allow a 2-$\sigma$ deviation from our average values, the discrepancies 
of ten pulsars can be explained by RISS.
For pulsars PSR B0919+06, PSR B1508+55 and PSR B1540$-$06, the discrepancies are much larger and 
an RISS explanation may not be adequate.
A partial explanation for pulsars PSR 0329+54 and PSR B1540$-$06 might be our instrumental limitations,
due to which the intensity patterns are barely resolved in frequency, thereby making the \nd values 
less accurate.

Evidence for another reason that can give rise to large discrepancies 
can be found in our data itself.
Interestingly, our observations show that average \nd values do not remain 
stable for pulsars PSR B0834+06, PSR B1133+16 and PSR B1919+21 from session to session, 
despite averaging over a time span
of about  100 days for each session, much larger than their RISS time scales.
Variations between the successive sessions of observations are found to be considerable $-$
in some cases as much  as factor of 2 to 3.
Thus our data show evidence for significant long-term variations of strengths of scattering on time scales 
$\sim$ years, something that has not been reported before.
It is not very clear what can cause such effects, but they are difficult to understand in terms of RISS
models based on simple power-law forms of density spectrum and a stationary statistics.
Whatever be the cause, such effects could explain the discrepancies seen in Table 6.
For example, PSR B1133+16 shows a systematic variation of \avnd from 434 kHz to 1434 kHz over a period 
$ \sim $ 900 days,
where \avnd from the first session is in agreement with values from literature. 
Similarly, for PSR B0834+06, \avnd values range from 350 kHz to 610 kHz, which 
is comparable to the differences between our data and the earlier measurements 
(except the value from Cordes et al. 1985).
Therefore, it is possible that the differences might be simply due to the measurements from 
observations made over separations of  years.

The frequency scaling of decorrelation bandwidth is highly sensitive to the nature of the density spectrum
assumed. 
The values listed, which are calculated for $ \alpha ~ = ~ { 11 \over 3 } $, will substantially differ 
if the spectrum is steeper.
The exact value of $ \alpha $ is still uncertain and there are conflicting 
interpretations from different scintillation experiments.
Recently, Armstrong et al. (1995) have shown that the overall density spectrum in
the nearby ISM ($ \la $ 1 kpc), extending over about 10 orders of magnitude of scale sizes, 
is closer to a Kolmogorov form,
but it is not clear if this is valid for all lines of sight.
It is possible that an incorrect frequency scaling has  also contributed, at least partially,
to the present discrepancies.

The situation with \td measurements is similar, though the disagreements are less pronounced.
Like the case with \nd values, RISS based on simple models fail to account for some of the 
observed discrepancies. 
It is possible that effects that can account for the discrepancies of \nd might explain 
the \td measurements as well.
In addition, an apparent change in the scintillation pattern speed, due to reasons such as earth's 
orbital motion (\vobs) and bulk motion of the medium (\virr), can also lead to changes in \td.
For example, in the case of PSR B2016+28, the variation  of scintillation time due to earth's motion
is substantial (Gupta et al. 1994), and therefore dependent on the epoch of observation.
Other pulsars where this effect is likely to be important are PSRs B1540$-$06 and B1604$-$00.

By and large, we find the discrepancies between our measurements (of both \nd and \td) and others 
to be more or less unbiased, except with those from Cordes et al. (1985).
But we also note that the measurements given in Cordes (1986) are from a more extensive data 
set and later than those reported in Cordes et al. (1985).
RISS seems to be the likely explanation for some of the discrepancies,
but there are several exceptions for which an RISS explanation is inadequate.
Long-term variations seen with \avnd (e.g. for PSR B1133+16) can also be interpreted as large-scale spatial
variations in \cn (say, over $ \sim $ 50$-$100 AU), and this could be, at least partly, responsible for some 
of the unexplained discrepancies.
Another possible of source of discrepancies is an incorrect scaling of the measurements with frequency.


\subsubsection{Flux Density}

We compare our flux density measurements (\Sort) (column 7 of Table 4) with the known values from the 
literature. 
Measurements at 400 MHz (\Slit) taken from Gould (1994) (hereinafter G94) (for 15 pulsars) and 
Taylor, Manchester \& Lyne (1993) (hereinafter TML93) are listed in columns (8) and (9) of Table 4. 
Our values, though less prone to errors due to RISS-induced fluctuations, may underestimate 
the true flux densities in the case of pulsars with a substantial fraction of linearly polarized radiation.
On the other hand, it is possible that measurements from G94 and TML93 are not averaged out for 
fluctuations due to RISS.
Our observations show large-amplitude variations in the flux density, where the individual measurements 
differ as much as by a factor 3$-$5.
The considerable discrepancies (as much a factor 2$-$3) seen between the values from G94 and TML93 are, 
therefore, explicable in terms of RISS.
Because of this, a detailed comparison between \Sort and \Slit may not be meaningful.

A first order comparison shows that, for 9 pulsars, the two values are comparable.
For PSRs B1540$-$06, B1604$-$00, B1747$-$46 and B2310+42, \Sort is comparable to \Slit from TML93.
For PSR B1133+16, \avSort $\approx$ \Slit from TML93 and for PSR B1919+21, \avfd from session II agrees 
with \Slit of TML93.
Similarly, the average value from sessions II$-$IV of PSR B0834+06 is comparable to \Slit of G94.
For PSR B0628$-$28, \Sort $\approx$ \avSlit (ie, the average of \Slit from G94 and TML93), and for 
PSR B2327$-$20, \Sort $\approx$ \Slit from G94.
For PSRs B0823+06, B0919+06 and B2045$-$16, our values are larger than \Slit by a factor $\sim$ 1.5$-$2.5,
whereas for PSRs B0329+54, B1508+55, B2016+28 and B2020+28, they are somewhat lower ($\sim$ 0.4$-$0.8 \Slit).
Two special cases are PSRs B1237+25 and B1929+10, for which \Sort is considerably lower; by $\sim$ 4 
times for the former and by $\sim$ 6$-$12 times for the latter.
Such a large discrepancy can be attributed to the fact that these pulsars show a large fraction of linearly
polarized radiation (with fractional linear polarizations (at 400 MHz) of $\approx$ 0.6 and $\approx$ 0.8 
respectively), which can make our values to underestimate the true values by a factor as much as $\sim$ 3$-$5.


\subsubsection{Scintillation Pattern Speeds and Proper Motion Speeds of Pulsars}

Scintillation speeds are considered to be good indicators of pulsar velocities and several authors 
have studied the correlation between the two (Lyne \& Smith 1982; Cordes 1986; Gupta 1995). 
In estimating scintillation speeds, different  authors use different values for the constant \Av in 
equation (9).
It was taken to be $ 2 \times 10^4 $ by Lyne \& Smith (1982), $ 1.27 \times 10^4 $ by Cordes (1986), 
while Gupta et al. (1994) suggest a value of $ 3.85 \times 10^4 $,
which has been adopted by us. 
Our estimates (column (5) of Table 4) are for a thin screen placed midway between the pulsar and the 
observer.
More refined estimates can be obtained using the method described in Cordes \& Rickett (1998) for
distributed scattering medium. 
It may be mentioned that the thin-screen method overestimates the scintillation speed compared to 
a uniformly distributed medium with a simple Kolmogorov form of density spectrum.

Proper motion measurements are available for 15 of our pulsars (Taylor et al. 1993, and references therein).
We have used them in combination with most recent distance estimates (Taylor et al. 1993) to estimate 
the pulsar velocities (\vprop) and these are given in column (6) of Table 4.
Values of \vprop are highly uncertain for 4 pulsars (PSRs B0628$-$28, B1604$-$00, B1919+21, B2016+28) 
as the error in the measurement is larger than the estimate of the proper motion itself.
Also, for PSRs B0919+06 and B1540$-$06, there are considerable uncertainties (60$-$70\%).
For the rest of the pulsars, the uncertainties in \vprop are \la 25\%.
These uncertainties are solely due to the measurement errors in proper motions. 
Taking into consideration the errors in distance estimates (typically 25\%), the actual errors 
in \vprop will be much larger for most pulsars.

Taking into consideration the uncertainties in \viss and \vprop, one can see that, for a number of pulsars, 
\viss is a reasonably good indicator of the proper motion speed.
A plot of \vprop vs \viss is shown in Fig. 6, where a significant correlation between the two can be seen
(correlation coefficient of 0.75).
The best agreement between \viss and \vprop is seen for PSR B1237+25.
The measurements are consistent for PSR B0919+06, but here the uncertainty in the proper motion is 
very large (70\%).
For PSRs B0834+06 and B1133+16, \viss estimates from part of the data agree with \vprop.
If we consider 3-$\sigma$ uncertainties in both \viss and \vprop, the two speeds are consistent 
for a fairly large number of measurements. 
Two exceptional cases are PSR B1508+55 with \viss $ \approx $ 0.5 \ \vprop, and PSR B2020+28 
with \viss $ \approx $ 2.5 \vprop.
For PSR B1508+55, the observed discrepancy is consistent with the estimated ``z-height'' of this pulsar and the
expected ``z-height'' of the electron density layer in the Galaxy, as discussed in detail in Gupta (1995) 
(hereinafter G95).
For PSR B2020+28, our observations confirm the discrepancy noted by G95.
The likely explanation for this is enhanced scattering from the Local Orion Arm of the Galaxy, 
as postulated in G95.
Significant discrepancies are also seen for \normb, \elevb and \elevc.

By and large, scintillation speeds can be used as rough indicators of proper motion speeds.
The discrepancies seen can result from the distance estimate used and/or an asymmetric location of 
the effective scattering screen (cf. G95).
Estimates of \viss are comparatively less prone to distance uncertainties, but critically depend on 
the relative location of the scattering screen with respect to the pulsar and the observer.
A detailed treatment of such discrepancies is beyond the scope of the present work.
The question of long-term stability of \viss also seems to have some relevance, as we find some evidence for
significant variation in \viss between different observing sessions for PSRs B0834+06, B1133+16 and B1919+21,
much in contradiction with the general expectations.
We discuss the issue of long-term ($ \sim $ year) stability of \nd, \td and \viss in the following section.


\subsection{Long-term Stability of Scintillation Properties and Pulsar Fluxes}

On the basis of results obtained from the observations of four pulsars 
for which we have multiple observing sessions $-$ PSR B0823+26,
PSR B0834+06, PSR B1133+16 and PSR B1919+21, we briefly discuss the issue of long-term stability
($ie.,$ over time scales much larger than RISS time scales) of scintillation properties and pulsar fluxes.
Observations were made over a fairly large number of epochs, spanning sufficiently long time spans
(66 to 120 days), except \eleva, where the statistics is poor in terms of number of measurements 
(\nep = 6) and observations span only 19 days.
The two sessions of PSRs B0823+26 and B1919+21 span 305 days and the data of PSRs B0834+06 and B1133+16
cover $ \approx $ 930 days.
This data allow us to examine the stability of measurements over time scales much larger than RISS time 
scales for these pulsars.

In general, we find the basic scintillation properties  \nd and \td
vary significantly between successive observing sessions.
Sometimes, considerable variations are seen in the average flux densities as well.
The observed long-term variations in the average values \avnd, \avtd, \avfd and \viss 
($ie.,$ computed using \ndg and \tdg, as described in \S 3.3.4) are summarized in Fig. 7.
The uncertainties shown by the vertical error bar are estimated as $ (\vmeas + \vstat)^{0.5} $.
The horizontal bar represents the time span of observation.

The changes in \avnd of PSR B0823+26 between the two sessions are within 
their statistical uncertainties, but the 18\% reduction in \avtd and the $ \sim $ 13\% increase in \viss
are larger than the uncertainties.
The increase in \viss can be explained by the 
transverse component of Earth's orbital motion (\vep) to the line-of-sight to
this pulsar, which increases from $ \approx $
$-1$ \velu in the first session to $ \approx $ 19 \velu in the second.
This naturally explains the apparent reduction in \avtd.
However, the 35\% reduction in \avfd from session I to II is not reconcilable within the
uncertainties (at $\pm$1-$\sigma$ level).

For PSR B0834+06, measurements of \avnd are stable (within the uncertainties) for the first three sessions, 
but there is $ \sim $ 75\% increase from III to IV. 
\avtd also shows a similar trend between these two sessions showing a  24\% increase, while  
\viss and \avfd remain stable within the measurement errors.
The variation of \avnd and \avtd between the sessions III and IV can be understood 
by a reduction of effective scattering strength (\intcn) by $ \sim $ 30\%.
The expected  \avtd of  340 secs ($ \td \propto \diff ^{-1} $), is consistent with 
the measured value $ 322 \pm 17 $ secs.
The reason for such a variation of \cn is, however, unclear.

Although \avnd remains more or less stable from session I to III for PSR B0834+06, 
there is a steady and significant reduction in \avtd, 
which is also reflected as an increase in \viss.
This pulsar is known to have a proper motion of $ 174 \pm 20 $ \velu, which is consistent with 
\viss measurements from the sessions I and II. 
The change in \avep from $ \approx $ 11 \velu to $ \approx $ 19 \velu can only partly 
account for the decrease in \avtd between the sessions I and II.
Similarly, \avep values of sessions I and IV, 11 \velu and $-$15 \velu, imply a relative change 
of $ \sim $ 26 \velu in \viss, which is about 50\% of the observed variation.
Thus the variations of \avtd are only partially explainable in terms of \vobs.
Further, there is a significant reduction (by 30\%) in \avfd from session I to II, which cannot
be reconciled to RISS modulations.

Long-term variations of scintillation characteristics of PSR B1133+16 are much more complex than the 
rest of the pulsars.
There is a factor of two increase in \avnd and \avtd between the sessions
I and II.
The variations in \avnd and \avtd are not consistent with a simple reduction in the scattering 
strength (\cn).
\avnd shows a remarkable increase between the sessions II and III, whilst \avtd and \avfd remain
more or less stable.
These variations are reflected in the estimated \viss values, which range from 
$ 335 \pm 10 $ \velu (in session II) to $ 490 \pm 38 $ \velu (in the initial session).
This pulsar also exhibits remarkable long-term changes in the dynamic spectra
where \nd ranges from  values as low as $ \sim $ 300 kHz 
(at some epochs of initial session) to as large as $ \sim $ 7 MHz (at some epochs of session III).
This is a unique pulsar which shows three different `modes' of scintillation $-$, 
(i) \nd $ \sim $ a few 100 kHz and \td $ \sim $ 100 secs (sessions I and second half of session II),
(ii) \nd $ \sim $ 1 MHz and \td $ \sim $ 200 secs (first half of session II), and 
(iii) \nd $ \sim $ several MHz and \td $ \sim $ 200 secs (session III).
The long-term variability of scintillation properties of this pulsar cannot be explained in terms of RISS,
variations in the strength of scattering (\cn) or the effect due to the Earth's orbital motion.
It may also be mentioned that this pulsar is fairly close-by (D $ \approx $ 270 pc) and is known to 
have shown multiple imaging events (Hewish et al. 1985; Cordes \& Wolszczan 1986; our observations).

For PSR B1919+21, substantial changes are seen in \avnd and \avtd, but with opposite trends.
This is also reflected as a remarkable ($ \sim $ 70\%) increase in \viss from session I to II.
This pulsar is known to have a proper motion \vu $ \sim $ 
$ 122 \pm 228 $ \velu, which will be consistent with \viss estimate from either of the sessions.
The increase in \avnd from 279 kHz to 510 kHz between the two sessions 
implies a reduction in the small-scale scattering 
(\intcn) by $ \sim $ 35\%, which would have resulted in \avtd $ \sim $ 555 secs if \viss $ \approx $
\vu for both the sessions.
The discrepancy in the two \viss values are not reconcilable within the uncertainties due to refractive 
modulations.
On examining the effect of the earth's orbital motion on \viss measurements of this pulsar, we find the 
relative change in \avep to be $ \approx $ 30 \velu (\avep values are $ \approx $ $-$5 \velu and 
$ \approx $ 25 \velu for the sessions I and II respectively).
The effect of observer's motion can, therefore, only partly account for the discrepancy.
The bulk flow of the density irregularities (\virr) is another possible explanation of discrepancy,
but the required \virr $ \sim $ 55 \velu is highly improbable.
Bondi et al. (1994), based on their one-year flux modulation studies of low frequency variables, argue
that \virr $ < $ 10 \velu.
Thus the observer's motion and the motion of the medium do not satisfactorily account for the observed 
variability of \viss values.
Another effect that can give rise to modulation of \viss is the angular wandering of the source 
position caused by the refractive effects, which we will discuss in a subsequent paper.
There is 42\% reduction in \avfd from session I to II, which is significantly larger than the
statistical fluctuations that can be expected due to RISS.

Systematic observations to study the long-term stability of basic scintillation properties 
(\nd and \td) and pulsar fluxes have not been attempted before.
The present observations show some evidence to suggest that these quantities do not remain stable  
for $some$ pulsars despite averaging over periods much longer than the RISS time scale.
Such a property is unexpected from the current RISS models.
Normal RISS is expected to give rise to fluctuations of scintillation observables 
over refractive time scales and no ISS effect is known to date which can cause long-term 
(session-to-session) variations seen in the present observations.
Since the measurements have been made with the same experimental set-up, 
an observational bias is highly unlikely.
The observed variations of \avnd and \avtd cannot be fully attributed to effective changes in scattering 
strength \cn and/or the effect due to earth's orbital motion around the Sun.
Substantial variations of \cn are not expected over the transverse extents of $ \sim $ 50$-$100 
AU that has been probed by the present data. 
The observed variability of the flux is also anomalous and there are no 
intrinsic or extrinsic effects known that can produce significant flux variations 
over time scales $ \sim $ 1$-$2 years.
All these conclusions are, however, based on  only $three$ pulsars 
and systematic observations of more pulsars are needed for understanding these effects better.


\section{Summary and Conclusions}

We have made an extensive series of scintillation observations for 18 pulsars in the DM range $3-35$ \dmu 
using the Ooty Radio Telescope at 327 MHz.
The dynamic scintillation spectra of these pulsars were monitored at $10-90$ epochs over time spans 
of 100 to 1000 days during $1993-95$.
In this paper, we have presented the basic results from these observations.
Some of the implications of these results, such as constraints on the spectrum of electron density
fluctuations in the ISM and implications for the theoretical models of refractive scintillation, 
are addressed in two subsequent papers.

Two-dimensional auto-covariance function (2-D ACF) were computed for the dynamic spectra, and are 
used to estimate the quantities such as decorrelation bandwidth (\nd), scintillation time scale (\td) 
and the drift rate of intensity patterns (\dtn) for each observation.
Time series of these quantities and flux density (F) are presented, and can be used to investigate 
various refractive effects in pulsar scintillation.
All the four quantities show large-amplitude fluctuations (as much as by a factor $3-5$) over time spans 
of about 100 days, and are also found to vary significantly over time scales as short as a few days.

Several pulsars show pronounced drifting of intensity patterns in dynamic spectra; 
of which PSRs B0834+06, B1133+16, B1919+21, B2045$-$16 form some of the best examples.
For many pulsars, the measured drift slopes (\dtn) show gradual variations with time, and undergo several 
sign reversals during the observing time spans. 
Data of PSRs B0823+26, B0919+06 and B2020+28 illustrate this property, which is in
accordance with the expectations of RISS theory.
However, there are some pulsars for which no sign reversals of the drift slopes are seen over many months. 
Data of PSRs B0834+06 and B1919+21 are best examples of such ``persistent drifts''.
Out of the four well separated observing sessions of PSR B0834+06 spanning $ \sim $ 1000 days, 
persistent drifts are observed in the first three, whereas the data from the final session show 
several occasions of sign reversals of pattern slopes.

PSR B1133+16, which was observed for 3 sessions spanning $ \sim $ 1000 days, shows remarkable changes in 
its dynamic spectrum from one session to another.
The data from the initial session are characterized by scintles that are fairly narrow in 
both time ($ \sim $ 100 sec) and in frequency ($ \sim $ 300 kHz).
The data taken one year after show much broader scintles ($ \sim $ 200 sec 
and $ \sim $ 1 MHz).
An episode of ``interstellar fringing'' is observed on 30 April 1994, which is followed by data characterized 
by narrower scintles (similar to the first session).
Data from the final session show patterns that decorrelate over a wider range of frequency 
($ \sim $ 2 MHz)  and time ($ \sim $ 200 sec).
Such dramatic variations are not seen for any other pulsar.
The pulsar PSR B1919+21 is found to show fine, deep intensity modulations 
over and above random intensity modulations in time and frequency caused by ISS.
These modulations occur over typical time scales of a minute (much shorter than DISS time scale) with level
of modulations as much as 50\%.
The modulations  are broadband and persist throughout our observations which 
suggests that they are intrinsic to the pulsar.

To obtain more reliable estimates of the average values of \nd, \td and \dtn from all the
observations of a given pulsar, we have computed the Global ACF (GACF) for each pulsar.
While the results from GACF and time series methods agree in general, there are considerable differences 
in some cases, specially when the statistics is poor.
Since the GACF method is expected to yield more robust estimates of the average properties, we prefer them 
over the time series averages.
Estimates of decorrelation bandwidth (\ndg) and scintillation time scale (\tdg) from GACFs are used to 
estimate parameters such as the line-of-sight averaged strength of scattering (\avcn), the strength of 
scattering parameter ($u$) and the scintillation pattern speed (\viss).
A comparison between the scintillation speeds derived from our measurements and proper motion speeds 
of pulsars show that scintillation speeds are reasonably good indicators of proper motion speeds.

The present observations have resulted in a significant improvement in terms of accuracy and reliability 
of the scintillation measurements compared to those available in the published literature.
For pulsars PSRs B$1747-46$, B2310+42, B$2327-20$, scintillation parameters are reported for the first time.
There are considerable differences between our measurements of decorrelation bandwidth and scintillation time
scale and various earlier measurements.
Though RISS seems to be a likely explanation for some of the differences, it does not satisfactorily 
account for all discrepancies.

There is some evidence from our data for fluctuations of scintillation properties and flux densities 
over time scales much longer than that of RISS. 
For pulsars studied over  multiple sessions, significant variations from session to session are sometimes seen 
in one or more of the parameters \avnd, \avtd and \avfd.
Long-term ($ \sim $ year) variations of this kind are not expected from current RISS models.
Some of the variations of \avtd can be accounted for by variations in scintillation pattern velocities 
due to the earth's orbital motion.
The \avnd variations indicate variations of strengths of scattering along certain lines-of-sight.
No obvious explanation can be found for the variations of \avfd.
Possible implications of these results are a non-stationary nature of the ISM over $ \sim $ 50$-$100 AU 
or some hitherto unidentified form of ISS.


{\it Acknowledgments:}
The authors would like to thank J. Chengalur and M. Vivekanand for a critical reading of the manuscript
and giving useful comments.
We thank V. Balasubramanian for the telescope time and technical help with the observations.
We also thank the staff of Radio Astronomy Centre, Ooty, for their assistance with the observations.
We thank our referee for comments, which helped us in clarifying a number of things related to the 
work presented in this paper.


\clearpage

\begin{appendix}

\section{Sources of Errors on the Average Scintillation Properties and Flux Densities}

Here we estimate the errors on the average values of scintillation properties (\ndg, \tdg and \dtng)
and flux density (\avfd).
These include (i) the measurement error (\smeas), and (ii) the statistical error due to the number of 
refractive cycles spanned by the measurements (\sstat).
The measurement error depends largely on the number of epochs of measurements (\nep).
For \ndg, \tdg and \dtng, which are estimated from GACF, \smeas 
has contributions mainly from 2 sources $-$ the error from the gaussian fit (as described in \S 3.3.1) 
to the GACF (denoted as \smodg and the estimation error due to finite number of scintles contributing 
to the GACF (denoted as \sestg).
To compute \sestg, we treat treat \ndg and \tdg to be representatives of a single dynamic spectrum of 
duration $ \sum _{i=1} ^{i=\nep} \tobsi $, where \tobsi is the duration of observation at $ i^{th} $ epoch.
The fractional estimation error (\sestg) is given by 

\begin{equation}
\sestg = \left[ f_d \left( { \sum _{i=1} ^{i=\nep} \bobsi \tobsi \over \ndg \tdg } \right) \right] ^{-0.5}
\end{equation}

\noindent
where \bobsi is the observing bandwidth used at $ i^{th} $ epoch and $f_d$ is the filling fraction of the number
of scintles, for which we assume a moderate value of 0.5.
The measurement error on the above 3 quantities is estimated as

\begin{equation}
\smeasg = \left[ \left( \smodg \right) ^2 + \left( \sestg \right) ^2 \right] ^{0.5}
\end{equation}

\noindent
The errors obtained in this manner are given in columns (3) and (4) of Table 5, for \ndg and \tdg respectively.
Typical values range from 2 to 5\%, with somewhat larger values for PSRs B1237+25 and B1929+10,
for which \smeasg $ \sim $ 10\%.

The measurement error in the flux density, \smeasf, is estimated as 

\begin{equation}
\smeasf = \left[ \left( { 1 \over \nep } \right) ^2 
\left( \sum _{i=1} ^{i=\nep} \left\{ \sfi \right\} ^2 \right) \right] ^{0.5}
\end{equation}

\noindent
where \sfi is the measurement error in flux density at $ i^{th} $ epoch, and it includes the errors due to 
calibration (\scal) and estimation error (\sest) due to finite number of scintles at that epoch.
We have adopted a conservative approach of 10\% uncertainty in calibration for all the epochs.
The column (5) of Table 5 gives \smeasf, typical values of which range from 2 to 5\%, except for 
PSRs B1237+25 and B1929+10, for which \smeasf $ \sim $ 8$-$9\%.

A direct estimation of the statistical error is not possible as our time series are not good enough
to estimate the time scale of fluctuations of the quantities.
However, we get a first order estimate based on the expected refractive time scale (\tref), which 
can be estimated from measurements of decorrelation bandwidth and scintillation time scale.
We estimate the statistical uncertainty as 

\begin{equation}
\sstat = \left( { x_{rms} \over \sqrt { \nref } } \right)
\end{equation}

\noindent
where \xrms denotes the rms estimate of the quantity $x$ under consideration, and obtained from the time
series.
A rough estimate of the number of refractive cycles of fluctuations (\nref) is given by \tsp/\tref, where
\tsp is the time span of observation.
The refractive time scale is taken as $ \tref \approx u^2 \td $ (Rickett 1990). 
We use values of \ndg and \tdg to get estimates of \tref.
The errors computed in this manner are given in columns (6), (7) and (8) of Table 5, 
for \avnd, \avtd and \avfd respectively.
Typical values range from 5 to 15\% for \avnd, 3 to 10\% for \avtd and 5 to 20\% for \avfd.

\end{appendix}


\clearpage


\begin{center} {\large\bf Figure Captions} \end{center}

\begin{description}

\item [Figure 1]
Sample dynamic spectra of selected pulsars shown as a gray-scale plot of intensity versus time and frequency.
The darker regions represent higher intensity values. 
The white regions correspond to 20\% of the mean intensity and the black regions to four times the mean 
intensity; in between, the intensity values are linearly represented by the gray-scale.
The average flux density at each epoch of observation is given at the upper left corner of each panel, 
and the observing day is indicated at the upper right corner.
(a) PSR B0823+26 at  9 epochs spanning 57 days during October$-$December 1993,
(b) PSR B0834+06 at 25 epochs spanning 110 days during January$-$May 1993,
(c) PSR B0834+06 at 23 epochs spanning 100 days during April$-$July 1995,
(d) PSR B0919+06 at  9 epochs spanning  35 days during March$-$June 1994,
(e) PSR B1133+16 at 16 epochs spanning  65 days during March$-$May 1994,
(f) PSR B1133+16 at 16 epochs spanning  92 days during April$-$July 1995,
(g) PSR B1919+21 at  9 epochs spanning  23 days during April$-$May 1993, and
(h) PSR B2045$-$16 at 9 epochs spanning 40 days during December 1993$-$January 1994. 

\item [Figure 2]
Contour plots of two-dimensional auto-covariance functions (2-D ACF) of dynamic spectra 
shown for three pulsars.
There are 20 contours over the range zero to unity; successive contours are separated 
by an interval 0.05, and dashed contours represent negative values.
The date of observation is indicated at the top of each panel. 
(a) PSR B0823+26 at  9 epochs of observations spanning 54 days during April$-$May 1993,
(b) PSR B0919+06 at  9 epochs of observations spanning 35 days during April$-$June 1994,
and
(c) PSR B1919+21 at 25 epochs of observations spanning 87 days during October 1993$-$January 1994. 

\item [Figure 3(a)$-$(m)]
Typical dynamic spectra of 13 pulsars shown to illustrate their general characteristics.
The name of the pulsar is given at the top of each panel. 
The average flux density is given at the upper left corner and the date of observation at the upper right 
corner of each panel. 
The nature of gray-scale representation is similar to Fig. 1, except for the intensity levels corresponding 
to the white and black regions.
The white and black correspond to 20\% and four times the mean intensity respectively 
for PSRs B0628$-$28, B1237+25, B1604$-$00, B1929+10, B2020+28 and B2327$-$20.
They correspond to 30\% and four times the mean for PSRs B1133+16, B1540$-$06 and B2016+28, 
30\% and thrice the mean for PSRs B0329+54 and B1747$-$46, and
40\% and thrice the mean for PSRs B1508+55 and B2310+42.

\item [Figure 4]
Time series of basic scintillation measurements: 
decorrelation bandwidth (\nd), scintillation time scale (\td), 
flux density (F) and drift rate of patterns (\dtn). 
The uncertainties in the measurements indicate $\pm$1-$\sigma$ error estimates. 
The name of the pulsar and session ID (wherever needed) are given in the topmost panel of each figure. 
The solid markers (at either ends of the panel) indicate the average estimates of each time series. 
The dashed line in the plot of \dtn corresponds to the zero drift slope.
The starting date corresponding to the day number 1 is given at the bottom of each figure (X-axis label). 
(a) PSR B0329+54, (b) PSR B0628$-$28, (c)$-$(d) data from the two observing sessions (I and II) of 
PSR B0823+26,
(e)$-$(h) data from the 4 observing sessions (I to IV) of PSR B0834+06,
(i) PSR B0919+06, (j)$-$(k) data of PSR B1133+16 for the observing sessions II and III, (l) PSR B1237+25,
(m) PSR B1508+55, (n) PSR B1540$-$06, (o) PSR B1604$-$00, (p) PSR B1747$-$46,
(q)$-$(r) data from the two observing sessions (I and II) of PSR B1919+21, (s) PSR B1929+10, (t) PSR B2016+28,
(u) PSR B2020+28, (v) PSR B2045$-$16, (w) PSR B2310+42, and (x) PSR B2327$-$20. 

\item [Figure 5]
The Global 2-D ACFs (GACF) of 18 pulsars (25 data sets).
The name of the pulsar is given at the top of each panel. 
For PSRs B0823+26, B0834+06, B1133+16 and B1919+21, there are multiple plots, which are GACFs computed for 
different observing sessions. 
The labels A and B represent the part of the data for which GACF is computed (see Table 3). 
GACFs represent the average scintillation properties.
The plots shown here illustrate the general characteristics observed with the dynamic spectra of various 
pulsars and also variations in scintillation properties from pulsar to pulsar.

\item [Figure 6]
The available proper motion speeds (\vprop) of 11 pulsars are plotted against their scintillation 
speeds (\viss) derived from the present scintillation measurements. 
The dashed line is of unity slope. 

\item [Figure 7]
Plots illustrating the long-term ($ \sim $ several months to year) variations in the average
scintillation properties (\avnd, \avtd and \viss) and pulsar flux densities (\avfd). 
The pulsar name and the quantity plotted are given at the top of each panel. 
The vertical bar is the total uncertainty due to measurement and statistical errors. 
The horizontal bar represents the length of observing time span. 
The measurements are normalized to unity for their lowest values.

\end{description}


\begin{thebibliography}{}

\bibitem{} Armstrong, J. W., Rickett, B. J. \& Spangler, S. R. 1995, ApJ, 443, 209
\bibitem{} Bhat, N. D. R., Gupta, Y. \& Rao, A. P. 1997,
in Proceedings of the IAU Colloquium No. 166 ``The Local Bubble and Beyond'',
eds. D. Breitschwerdt, M. J. Freyberg, J. Tr\"umper, 
Lecture Notes in Physics 506, 211
(Springer-Verlag)
\bibitem{} Bhat, N. D. R., Gupta, Y. \& Rao, A. P. 1998, ApJ, 500, 262
\bibitem{} Blandford, R. D. \& Narayan, R. 1985, MNRAS, 213, 591
\bibitem{} Bondi, M., Padrielli, L., Gregorini, L., Mantovani, F., Shapirovskaya, N. \& Spangler, S. R. 1994, 
A\&A, 287, 390
\bibitem{} Cognard, I., Bourgois, G., Lestrade, J., Biraud, F., Aubry, D., Darchy, B. \& Drouhin, J. 1993,
Nature, 366, 320
\bibitem{} Cole, T.W., Hesse, H. K. \& Page, C. G. 1970, Nature, 225, 712
\bibitem{} Cordes, J. M. 1986, ApJ, 311, 183
\bibitem{} Cordes, J. M., Pidwerbetsky, A. \& Lovelace, R. V. E. 1986, ApJ, 310, 737
\bibitem{} Cordes, J. M. \& Rickett, B. J. 1998, Submitted 
\bibitem{} Cordes, J. M., Spangler, S. R., Weisberg, J. M. \& Clifton, T. R. 1988,
in AIP Conf. Proc. No. 174 - Radiowave Scattering in the Interstellar Medium,
ed. Cordes, J. M., Rickett, B. J. \& Backer, D. C.
(New York: AIP), 180
\bibitem{} Cordes, J. M., Weisberg, J. M. \& Boriakoff, V. 1985, ApJ, 288, 221
\bibitem{} Fiedler, R. L., Dennison, B., Johnston, K. J. \& Hewish, A. 1987, Nature, 326, 675
\bibitem{} Gould, D. M. 1994, Ph.D. thesis, University of Manchester
\bibitem{} Gupta, Y. 1995, ApJ, 451, 717
\bibitem{} Gupta, Y., Rickett, B. J. \& Coles, W. A. 1993, ApJ, 403, 183
\bibitem{} Gupta, Y., Rickett, B. J. \& Lyne, A. G. 1994, MNRAS, 269, 1035
\bibitem{} Gwinn, C. R., Taylor, J. H., Weisberg, J. M. \& Rawlings, L. A. 1986, AJ, 91, 338
\bibitem{} Hewish, A. 1980, MNRAS, 192, 799
\bibitem{} Hewish, A., Wolszczan, A. \& Graham, D. A. 1985, MNRAS, 213, 167
\bibitem{} Huguenin, G. R. \& Taylor, J. H. \& Helfand, D. J. 1973, ApJ, 181, L139
\bibitem{} Kaspi, V. M. \& Stinebring, D. R. 1992, ApJ, 392, 530 
\bibitem{} LaBrecque, D. R., Rankin, J. M. \& Cordes, J. M. 1994, AJ, 108, 1854
\bibitem{} Lestrade, J., Cognard, I. \& Biraud, F. 1995, 
in Millisecond Pulsars: A Decade of Surprise,
ed. Fruchter, A., Tavani, M. \& Backer, D. 
(San Francisco: Astronomical Society of the Pacific Conference Series), 357
\bibitem{} Lestrade, J., Rickett, B. J. \& Cognard, I. 1998, A\&A, 334, 1068
\bibitem{} Lyne, A. G. \& Smith, F. G. 1982, Nature, 298, 825
\bibitem{} Rickett, B. J. 1969, Nature, 221, 158
\bibitem{} Rickett, B. J. 1977, ARA\&A, 15, 479
\bibitem{} Rickett, B. J. 1986, ApJ, 307, 564
\bibitem{} Rickett, B. J. 1990, ARA\&A, 28, 561
\bibitem{} Rickett, B. J., Coles, W. A. \& Bourgois G. 1984, A\&A, 134, 390
\bibitem{} Roberts, J. A. \& Ables, J. G. 1982, MNRAS, 201, 1119
\bibitem{} Romani, R. W., Narayan, R. \& Blandford, R. D. 1986, MNRAS, 220, 19
\bibitem{} Scheuer, P. A. G. 1968, Nature, 218, 920
\bibitem{} Selvanayagam, A. J., Praveenkumar, A., Nandagopal, D. \& Veluswamy, T. 1993, 
IETE Technical Review, 10, 333
\bibitem{} Sieber, W. 1982, A\&A, 113, 311
\bibitem{} Smith, F. G. \& Wright, N. C. 1985, MNRAS, 214, 97
\bibitem{} Spangler, S. R. 1988,
in AIP Conf. Proc. No. 174 - Radiowave Scattering in the Interstellar Medium,
ed. Cordes, J. M., Rickett, B. J. \& Backer, D. C.
(New York: AIP), 32
\bibitem{} Stinebring, D. R. \& Condon, J. J. 1990, ApJ, 352, 207
\bibitem{} Stinebring, D. R., Faison, M. D. \& McKinnon, M. M. 1996, ApJ, 460, 460
\bibitem{} Subramanian, R. 1989, Ph.D. thesis, Indian Institute of Science
\bibitem{} Swarup, G., et al. 1971, Nature Phys Sci, 230, 185
\bibitem{} Taylor, J. H. \& Cordes, J. M. 1993, ApJ, 411, 674
\bibitem{} Taylor, J. H., Manchester, R. N. \& Lyne, A. G. 1993, ApJSS, 88, 529
\bibitem{} Wolszczan, A. \& Cordes, J. M. 1987, ApJ, 320, L35

\end{thebibliography}
\end{document}



\begin{deluxetable}{rlrllrllr}
\tablenum{1}
\tablewidth{0pt}
\tablecaption{The pulsar sample and the observing parameters}
\tablehead{
No. & Pulsar & DM & D &  Period of &  $ {\rm N_{ep} } $ & 
$ {\rm \Delta f _{ch} } $ & $ {\rm \Delta t } $ &
$ \tau _{r,exp} $ $ ^{\ddagger} $ \\
& & (${\rm{pc~cm^3}}$) & (pc)& observation & & (kHz) & (secs) & (days) \\ \\
(1) & (2) & (3) & (4) & (5) & (6) & (7) & (8) & (9) 
}
\startdata
 1 & PSR \ B$0329+54$ & 26.8 & 1430   &  Apr 95$-$Jul 95 &  14 &$140.6$&$7.15$       & 39.7 \\
 2 & PSR \ B$0628-28$ & 34.4 & 2140   &  Oct 93$-$Jan 94    &  17 &$140.6$&$12.44$   & 14.6 \\
 3 & PSR \ B$0823+26$ & 19.5 &$~380^a$ &Mar 93$-$Jan 94&  31 &$140.6^b$&$13.27$       &  5.5 \\
 4 & PSR \ B$0834+06$ & 12.9 & ~720   &Jan 93$-$Jul 95 &  93 &$140.6^b$&$12.74$      &  5.2 \\
 5 & PSR \ B$0919+06$ & 27.2 & 2970$^d$   &  Mar 94$-$Jun 94 &  19 &$140.6$&$21.53$      & 12.6 \\
 6 & PSR \ B$1133+16$ &  4.8 & ~270   &Feb 93$-$Jul 95&  59 &$140.6^b$&$11.88$       &  1.4 \\
 7 & PSR \ B$1237+25$ &  9.3 & ~560   &  Oct 93$-$Jan 94    &   9 &$140.6$&$13.82$   &  2.2 \\
 8 & PSR \ B$1508+55$ & 19.6 & 1930   &  Apr 95$-$Jul 95 &   9 &$140.6$&$14.79^c$    & 15.1 \\
 9 & PSR \ B$1540-06$ & 18.5 & 1160   &  Apr 95$-$Jul 95 &  12 &$140.6$&$14.18$      & 12.0 \\
10 & PSR \ B$1604-00$ & 10.7 & ~590   &  Apr 95$-$Jul 95 &  10 &$140.6$&$21.09$      & 13.4 \\
11 & PSR \ B$1747-46^{\dagger} $ & 21.7 & 1080   &  Apr 95$-$Jul 95 &  12 &$140.6$&$14.85^c$    &      \\
12 & PSR \ B$1919+21$ & 12.4 & ~660   &Mar 93$-$Jan 94&  63 &$140.6^b$&$13.37$       &  6.5 \\
13 & PSR \ B$1929+10$ &  3.2 & ~170   &  Mar 94$-$Jun 94 &   9 &$140.6$&$11.33$      &  2.7 \\
14 & PSR \ B$2016+28$ & 14.2 & 1100   &  Oct 93$-$Jan 94    &  20 &$140.6$&$13.95$   & 77.4 \\
15 & PSR \ B$2020+28$ & 24.6 & 1300   &  Mar 94$-$Jun 94 &  15 &$140.6$&$17.20^c$    & 22.5 \\
16 & PSR \ B$2045-16$ & 11.5 & ~640   &  Oct 93$-$Jan 94    &  35 &$140.6$&$19.62$   &  1.4 \\
17 & PSR \ B$2310+42^{\dagger} $ & 17.3 & ~960   &  Apr 95$-$Jul 95 &  10 &$140.6$&$17.47$      &      \\
18 & PSR \ B$2327-20^{\dagger} $ &  8.4 & ~490   &  Mar 94$-$Jun 94 &  18 &$140.6$&$32.87$      &      \\
\tablenotetext{\ddagger} {Estimates are based on the measurements from Cordes (1986).}
\tablenotetext{\dagger} {Scintillation measurements are not reported so far for these pulsars.}
\tablenotetext{a} {The distance estimates are from parallax measurements (Gwinn et al. 1986).}
\tablenotetext{b} {Part of the data were taken with $ \Delta f _{ch} $ = 281.20 kHz.}
\tablenotetext{c} {Part of the data were taken with larger $ \Delta t $ (twice the value given here).}
\tablenotetext{d} {Lower limit on the distance estimate.}
\enddata
\end{deluxetable}

\begin{deluxetable}{cccccccc}
\tablenum{2}
\tablewidth{0pt}
\tablecaption{Pulsars observed for multiple sessions}
\tablehead
{
  &              &   &    &    &    &    &     \\
No. & Pulsar & $ {\rm {N_{sess}} } $ &
\multicolumn{4}{c}{Number of epochs of observation per session (${\rm N_{ep}}$)} & $ {\rm T_{sp,max} } $ \\ \\
 & & & Session I & Session II & Session III & Session IV &  (days) \\
 & & & (Jan$-$May 93) & (Oct 93$-$Jan 94) & (Feb$-$Jun 94) & (Apr$-$Jul 95)  & \\
  &              &   &    &    &    &    &     \\
  (1) & (2) & (3) & (4) & (5) & (6) & (7) & (8)
}
\startdata
  &              &   &    &    &    &    &     \\
1 & PSR B0823+26 & 2 & 12 (66) & 21 (87) &    $ \cdots $  &    $ \cdots $  & 305 \\
  &              &   &    &    &    &    &     \\
2 & PSR B0834+06 & 2 & 27 (120) & 18 (74) & 21 (78) & 27 (100) & 935 \\
  &              &   &    &    &    &    &     \\
3 & PSR B1133+16 & 3 &  6 (19) &    $ \cdots $  & 25 (90) & 27 (95) & 928 \\
  &              &   &    &    &    &    &     \\
4 & PSR B1919+21 & 4 & 15 (59) & 48 (87) &    $ \cdots $  &    $ \cdots $  & 305 \\
  &              &   &    &    &    &    &     \\
\enddata
\end{deluxetable}

\newcommand{\dtnu}{${\rm sec~kHz ^{-1}}$}

\addtolength{\oddsidemargin}{-2.5cm}

\begin{deluxetable}{lccccccccc}
\tablenum{3}
\tablewidth{0pt}
\tablecaption{Measurements of Average scintillation parameters}
\tablehead
{
                       & \multicolumn{3}{c}{Time Series} & \multicolumn{3}{c}{Global ACF} & \multicolumn{3}{c}{Obs. Paras.} \\
Pulsar (Session) & 
$ \langle \nu _d \rangle $ & $ \langle \tau _d \rangle $ & $ \langle d t / d \nu \rangle $ &
$ \nu _{d,g} $ & $ \tau _{d,g} $ & $ ( d t / d \nu ) _g $ &
$ \Delta f _{ch} $ & $ \Delta t $ & $ {\rm N^{'} _{ep} } $ \\
 & (kHz) & (secs) & (\dtnu) &
 (kHz) & (secs) & (\dtnu) &
 (kHz) & (secs) & \\
(1) & (2) & (3) & (4) & (5) & (6) & (7) & (8) & (9) & (10) \\
}
\startdata
 PSR B$0329+54$     &~130$ \pm $~4&274$ \pm $~9& $ -0.09 \pm 0.04 $ &~165$ \pm $~~8& 307$ \pm $~19& $-0.08 \pm 0.25$ & 140 & 07.15 & 15 \\
 PSR B$0628-28$     &~221$ \pm $~8&469$ \pm $22& $ +0.04 \pm 0.08 $ &~203$ \pm $~11& 456$ \pm $~26& $+0.09 \pm 0.29$ & 140 & 12.44 & 16 \\
 PSR B$0823+26$(IA) &~244$ \pm $15&150$ \pm $12& $ +0.11 \pm 0.03 $ &~304$ \pm $~12& 145$ \pm $~~6& $+0.11 \pm 0.04$ & 280 & 13.27 & 04 \\
 PSR B$0823+26$(IB) &~287$ \pm $18&160$ \pm $11& $ -0.12 \pm 0.03 $ &~314$ \pm $~14& 162$ \pm $~~7& $-0.07 \pm 0.06$ & 140 & 13.27 & 06 \\
 PSR B$0823+26$(II) &~250$ \pm $~8&127$ \pm $~4& $ -0.06 \pm 0.02 $ &~261$ \pm $~10& 126$ \pm $~~5& $-0.05 \pm 0.05$ & 140 & 13.27 & 19 \\
 PSR B$0834+06$(IA) &~419$ \pm $14&413$ \pm $14& $ +0.41 \pm 0.02 $ &~454$ \pm $~11& 391$ \pm $~11& $+0.39 \pm 0.04$ & 280 & 12.74 & 17 \\
 PSR B$0834+06$(IB) &~493$ \pm $22&381$ \pm $16& $ +0.38 \pm 0.02 $ &~481$ \pm $~13& 353$ \pm $~10& $+0.33 \pm 0.05$ & 140 & 12.74 & 10 \\
 PSR B$0834+06$(II) &~369$ \pm $11&314$ \pm $11& $ +0.39 \pm 0.02 $ &~375$ \pm $~~7& 328$ \pm $~~7& $+0.37 \pm 0.04$ & 140 & 12.74 & 18 \\
 PSR B$0834+06$(III)&~353$ \pm $~8&259$ \pm $~7& $ +0.29 \pm 0.01 $ &~369$ \pm $~~4& 255$ \pm $~~3& $+0.28 \pm 0.02$ & 140 & 12.74 & 21 \\
 PSR B$0834+06$(IV) &~616$ \pm $17&321$ \pm $10& $ +0.08 \pm 0.01 $ &~599$ \pm $~~7& 309$ \pm $~~4& $+0.08 \pm 0.02$ & 140 & 12.74 & 27 \\
 PSR B$0919+06$     &~235$ \pm $~6&154$ \pm $~4& $ +0.01 \pm 0.02 $ &~256$ \pm $~~6& 160$ \pm $~~4& $-0.01 \pm 0.04$ & 140 & 21.53 & 18 \\
 PSR B$1133+16$(I)  &~434$ \pm $18&~89$ \pm $~4& $ +0.01 \pm 0.01 $ &~543$ \pm $~24& ~92$ \pm $~~4& $+0.02 \pm 0.03$ & 280 & 11.88 & 06 \\
 PSR B$1133+16$(II) &~838$ \pm $21&169$ \pm $~4& $ +0.03 \pm 0.01 $ &~816$ \pm $~11& 165$ \pm $~~2& $+0.05 \pm 0.01$ & 140 & 11.88 & 24 \\
 PSR B$1133+16$(III)&1399$ \pm $33&189$ \pm $~4& $ -0.05 \pm 0.01 $ &1301$ \pm $~12& 192$ \pm $~~2& $-0.05 \pm 0.01$ & 140 & 11.88 & 28 \\
 PSR B$1237+25$     &1146$ \pm $60&304$ \pm $16& $ +0.04 \pm 0.01 $ &1828$ \pm $174& 439$ \pm $~40& $-0.00 \pm 0.06$ & 140 & 13.82 & 09 \\
 PSR B$1508+55$(A)  &~208$ \pm $~9&139$ \pm $~6& $ +0.03 \pm 0.03 $ &~226$ \pm $~~6& 164$ \pm $~~5& $+0.03 \pm 0.05$ & 140 & 29.59 & 04 \\
 PSR B$1508+55$(B)  &~148$ \pm $~8&145$ \pm $~9& $ -0.02 \pm 0.06 $ &~168$ \pm $~15& 158$ \pm $~16& $-0.04 \pm 0.21$ & 140 & 14.79 & 04 \\
 PSR B$1540-06$     &~~90$ \pm $~3&492$ \pm $20& $ +0.22 \pm 0.12 $ &~111$ \pm $~~5& 526$ \pm $~25& $+0.22 \pm 0.36$ & 140 & 14.18 & 12 \\
 PSR B$1604-00$     &~376$ \pm $15&881$ \pm $38& $ +0.63 \pm 0.08 $ &~379$ \pm $~11& 933$ \pm $~30& $+0.34 \pm 0.19$ & 140 & 21.09 & 10 \\
 PSR B$1747-46$(A)  &~160$ \pm $~7&230$ \pm $11& $ +0.01 \pm 0.06 $ &~171$ \pm $~~7& 239$ \pm $~10& $+0.02 \pm 0.14$ & 140 & 29.69 & 07 \\
 PSR B$1747-46$(B)  &~130$ \pm $~7&202$ \pm $11& $ +0.09 \pm 0.07 $ &~158$ \pm $~~8& 215$ \pm $~11& $+0.10 \pm 0.14$ & 140 & 14.85 & 05 \\
 PSR B$1919+21$(IA) &~295$ \pm $17&454$ \pm $25& $ +0.14 \pm 0.05 $ &~338$ \pm $~11& 437$ \pm $~14& $+0.19 \pm 0.08$ & 280 & 13.37 & 06 \\
 PSR B$1919+21$(IB) &~269$ \pm $~9&382$ \pm $14& $ +0.55 \pm 0.04 $ &~285$ \pm $~~6& 374$ \pm $~~9& $+0.56 \pm 0.06$ & 140 & 13.37 & 09 \\
 PSR B$1919+21$(II) &~547$ \pm $11&338$ \pm $~8& $ +0.17 \pm 0.01 $ &~581$ \pm $~~9& 339$ \pm $~~5& $+0.13 \pm 0.02$ & 140 & 13.37 & 40 \\
 PSR B$1929+10$     &1199$ \pm $83&354$ \pm $24& $ +0.08 \pm 0.01 $ &1293$ \pm $~74& 348$ \pm $~19& $-0.01 \pm 0.04$ & 140 & 11.33 & 09 \\
 PSR B$2016+28$     &~201$ \pm $~8&999$ \pm $41& $ +0.30 \pm 0.14 $ &~206$ \pm $~19& 995$ \pm $100& $+0.52 \pm 1.07$ & 140 & 13.95 & 19 \\
 PSR B$2020+28$     &~199$ \pm $~5&271$ \pm $~8& $ -0.10 \pm 0.03 $ &~270$ \pm $~12& 279$ \pm $~12& $-0.17 \pm 0.13$ & 140 & 17.20 & 15 \\
 PSR B$2045-16$     &~576$ \pm $13&133$ \pm $~3& $ +0.06 \pm 0.01 $ &~539$ \pm $~17& 138$ \pm $~~4& $+0.05 \pm 0.02$ & 140 & 19.62 & 29 \\
 PSR B$2310+42$     &~~66$ \pm $~3&315$ \pm $13& $ -0.18 \pm 0.09 $ &~114$ \pm $~~3& 309$ \pm $~~9& $-0.12 \pm 0.14$ & 140 & 17.47 & 10 \\
 PSR B$2327-20$     &~248$ \pm $~8&514$ \pm $18& $ -0.04 \pm 0.07 $ &~268$ \pm $~18& 432$ \pm $~31& $+0.06 \pm 0.32$ & 140 & 32.90 & 18 \\
\enddata
\end{deluxetable}

\addtolength{\oddsidemargin}{+2.0cm}

\newcommand{\viss}{\mbox{${\rm V_{iss}} ~ $}}
\newcommand{\vprop}{\mbox{${\rm V_{prop}} ~ $}}
\newcommand{\velu}{\mbox{$ {\rm km ~ secs ^ {-1} } $}}
\newcommand{\cn}{\mbox{${\rm \overline { C_n^2 } } $}}

\begin{deluxetable}{clccccccc}
\tablenum{4}
\tablewidth{0pt}
\tablecaption{Derived Scattering Parameters and Pulsar Flux Densities}
\tablehead
{
No. & Pulsar & log \cn & $u$ & \viss & \vprop & $ S_{327} $ & \multicolumn{2}{c}{$ S_{400} $} \\
    &        &         &     &       &        &             & G94$^{\dagger}$ & TML93$^{\ddagger}$ \\
    &      &         &     & (\velu) & (\velu) & (mJy)    & (mJy) & (mJy)                   \\ \\
    (1) & (2) & (3) & (4) & (5) & (6) & (7) & (8) & (9)
}
\startdata
01 & PSR B$0329+54$     &$-4.11$ $ \pm $ 0.019 & 63.0 $ \pm $ 1.7 & 186 $ \pm $ 17 & 145 $ \pm $ ~10 &1230 $ \pm $ 38 & 1565 & 1650 \\
02 & PSR B$0628-28$     &$-4.51$ $ \pm $ 0.020 & 56.8 $ \pm $ 1.6 & 170 $ \pm $ 15 & 180 $ \pm $ 321 &~307 $ \pm $ ~9 & ~422 & ~200 \\
03 & PSR B$0823+26$(I)  &$-3.28$ $ \pm $ 0.012 & 46.0 $ \pm $ 0.8 & 264 $ \pm $ 14 & 196 $ \pm $ ~~7 &~178 $ \pm $ ~6 & ~~78 & ~~65 \\
04 & PSR B$0823+26$(II) &$-3.22$ $ \pm $ 0.015 & 50.1 $ \pm $ 1.0 & 294 $ \pm $ 18 &                 &~115 $ \pm $ ~3 &      &      \\
05 & PSR B$0834+06$(I)  &$-3.94$ $ \pm $ 0.010 & 37.4 $ \pm $ 0.5 & 184 $ \pm $ ~8 & 174 $ \pm $ ~20 &~250 $ \pm $ ~7 & ~146 & ~~85 \\
06 & PSR B$0834+06$(II) &$-3.86$ $ \pm $ 0.010 & 41.8 $ \pm $ 0.6 & 187 $ \pm $ ~8 &                 &~173 $ \pm $ ~5 &      &      \\
07 & PSR B$0834+06$(III)&$-3.85$ $ \pm $ 0.007 & 42.3 $ \pm $ 0.4 & 237 $ \pm $ ~7 &                 &~147 $ \pm $ ~4 &      &      \\
08 & PSR B$0834+06$(IV) &$-4.03$ $ \pm $ 0.008 & 33.0 $ \pm $ 0.4 & 250 $ \pm $ ~8 &                 &~150 $ \pm $ ~4 &      &      \\
09 & PSR B$0919+06$     &$-4.85$ $ \pm $ 0.010 & 50.5 $ \pm $ 0.7 & 642 $ \pm $ 26 & 919 $ \pm $ 662 &~~81 $ \pm $ ~2 & ~~55 & ~~60 \\
10 & PSR B$1133+16$(I)  &$-3.22$ $ \pm $ 0.018 & 34.7 $ \pm $ 0.9 & 490 $ \pm $ 38 & 475 $ \pm $ ~~8 &~315 $ \pm $ 14 & ~~95 & ~300 \\
11 & PSR B$1133+16$(II) &$-3.36$ $ \pm $ 0.008 & 28.3 $ \pm $ 0.3 & 335 $ \pm $ 10 &                 &~260 $ \pm $ ~7 &      &      \\
12 & PSR B$1133+16$(III)&$-3.53$ $ \pm $ 0.007 & 22.4 $ \pm $ 0.2 & 363 $ \pm $ 10 &                 &~300 $ \pm $ ~7 &      &      \\
13 & PSR B$1237+25$     &$-4.24$ $ \pm $ 0.043 & 18.9 $ \pm $ 1.1 & 271 $ \pm $ 47 & 303 $ \pm $ ~13 &~~29 $ \pm $ ~1 & ~135 & ~110 \\
14 & PSR B$1508+55$     &$-4.41$ $ \pm $ 0.018 & 57.6 $ \pm $ 1.4 & 451 $ \pm $ 35 & 913 $ \pm $ ~46 &~~60 $ \pm $ ~2 & ~105 & ~~95 \\
15 & PSR B$1540-06$     &$-3.80$ $ \pm $ 0.017 & 76.8 $ \pm $ 1.8 & ~80 $ \pm $ ~6 & 355 $ \pm $ 222 &~~34 $ \pm $ ~1 & ~~19 & ~~30 \\
16 & PSR B$1604-00$     &$-3.71$ $ \pm $ 0.018 & 41.5 $ \pm $ 1.0 & ~60 $ \pm $ ~5 & ~20 $ \pm $ ~47 &~~55 $ \pm $ ~2 & ~~98 & ~~60 \\
17 & PSR B$1747-46$     &$-3.89$ $ \pm $ 0.012 & 63.1 $ \pm $ 1.1 & 218 $ \pm $ 11 &                 &~~73 $ \pm $ ~2 &      & ~~70 \\
18 & PSR B$1919+21$(I)  &$-3.73$ $ \pm $ 0.011 & 45.9 $ \pm $ 0.7 & 132 $ \pm $ ~6 & 122 $ \pm $ 228 &~355 $ \pm $ 12 & ~~72 & ~200 \\
19 & PSR B$1919+21$(II) &$-3.95$ $ \pm $ 0.007 & 33.6 $ \pm $ 0.3 & 215 $ \pm $ ~7 &                 &~206 $ \pm $ ~4 &      &      \\
20 & PSR B$1929+10$     &$-3.16$ $ \pm $ 0.029 & 22.5 $ \pm $ 0.9 & 159 $ \pm $ 19 & ~86 $ \pm $ ~~6 &~~38 $ \pm $ ~3 & ~450 & ~250 \\
21 & PSR B$2016+28$     &$-3.98$ $ \pm $ 0.035 & 56.3 $ \pm $ 2.7 & ~56 $ \pm $ ~9 & ~12 $ \pm $ ~19 &~211 $ \pm $ ~8 &      & ~320 \\
22 & PSR B$2020+28$     &$-4.21$ $ \pm $ 0.017 & 49.2 $ \pm $ 1.1 & 250 $ \pm $ 18 & ~97 $ \pm $ ~22 &~~46 $ \pm $ ~1 &      & ~110 \\
23 & PSR B$2045-16$     &$-3.90$ $ \pm $ 0.012 & 34.8 $ \pm $ 0.6 & 501 $ \pm $ 25 & 289 $ \pm $ ~71 &~226 $ \pm $ ~4 & ~~87 & ~125 \\
24 & PSR B$2310+42$     &$-3.66$ $ \pm $ 0.011 & 75.7 $ \pm $ 1.2 & 126 $ \pm $ ~6 &                 &~108 $ \pm $ ~4 & ~~49 & ~~95 \\
25 & PSR B$2327-20$     &$-3.43$ $ \pm $ 0.025 & 49.4 $ \pm $ 1.7 & ~99 $ \pm $ 11 &                 &~~31 $ \pm $ ~1 & ~~36 & ~~50 \\
\tablenotetext{\dagger}{Measurements at 400 MHz from Gould (1994)}
\tablenotetext{\ddagger}{Measurements at 400 MHz from Taylor, Manchester \& Lyne (1993)}
\enddata
\end{deluxetable}

\begin{deluxetable}{clcccccc}
\tablenum{5}
\tablewidth{0pt}
\tablecaption
{Error estimates on Scintillation measurements and Pulsar flux densities}
\tablehead
{
No. & \multicolumn{1}{c}{Pulsar (Session)} & 
\multicolumn{3}{c}{Measurement Errors in \%} &
\multicolumn{3}{c}{Statistical Uncertainties in \%} \\
        &       & \multicolumn{3}{c}{ $ \sigma _{meas} $ } & 
\multicolumn{3}{c}{ $ \sigma _{stat} $ } \\
        &       & 
($ {\nu _d} $) & ($ {\tau _d} $) & ( {F} ) &
($ {\nu _d} $) & ($ {\tau _d} $) & ( {F} ) \\
        &       &       &       &       &       &       &       \\
(1) & (2) & (3) & (4) & (5) & (6) & (7) & (8)
}
\startdata
01& PSR B$0329+54$	&  5.3 & 6.3 & 3.4 & 18.3 &  7.8 & 10.6 \\
02& PSR B$0628-28$	&  5.6 & 5.9 & 3.2 & 20.7 & 10.3 & 13.5 \\
03& PSR B$0823+26$(I)	&  3.5 & 3.5 & 3.7 &  8.9 &  4.2 & 10.1 \\
04& PSR B$0823+26$(II)	&  4.1 & 4.2 & 2.7 &  7.1 &  2.4 &  7.3 \\
05& PSR B$0834+06$(I)	&  2.7 & 2.8 & 3.0 &  8.0 &  6.0 & 12.0 \\
06& PSR B$0834+06$(II)	&  2.7 & 2.8 & 3.2 & 10.7 &  4.7 & 10.7 \\
07& PSR B$0834+06$(III)	&  1.8 & 1.9 & 2.7 &  5.4 &  3.3 &  9.2 \\
08& PSR B$0834+06$(IV)	&  2.1 & 2.1 & 2.9 &  7.5 &  4.9 &  6.9 \\
09& PSR B$0919+06$	&  2.7 & 2.7 & 2.9 &  6.7 &  4.9 & 12.6 \\
10& PSR B$1133+16$(I)	&  5.0 & 5.3 & 4.5 &  7.1 &  5.1 &  5.4 \\
11& PSR B$1133+16$(II)	&  2.1 & 2.0 & 2.5 &  5.6 &  4.0 &  3.1 \\
12& PSR B$1133+16$(III)	&  1.9 & 1.9 & 2.6 &  4.4 &  1.9 &  3.7 \\
13& PSR B$1237+25$	& 11.8 &11.6 & 8.1 &  5.0 &  4.7 & 10.5 \\
14& PSR B$1508+55$	&  4.9 & 5.4 & 4.1 & 10.3 &  5.1 &  8.2 \\
15& PSR B$1540-06$	&  4.7 & 5.1 & 3.8 & 16.2 & 12.5 & 30.6 \\
16& PSR B$1604-00$	&  4.9 & 5.1 & 5.5 & 19.6 &  7.6 & 20.9 \\
17& PSR B$1747-46$	&  3.4 & 3.5 & 3.4 & 13.1 &  5.7 &  8.1 \\
18& PSR B$1919+21$(I)	&  3.0 & 3.1 & 3.5 &  6.9 &  8.5 & 19.8 \\
19& PSR B$1919+21$(II)	&  2.1 & 2.1 & 2.2 &  8.0 &  4.5 &  9.8 \\
20& PSR B$1929+10$	&  7.9 & 7.7 & 9.1 &  8.4 &  3.5 & 14.4 \\
21& PSR B$2016+28$	&  9.6 &10.4 & 4.1 & 31.0 & 22.4 & 26.4 \\
22& PSR B$2020+28$	&  4.6 & 4.7 & 3.6 &  6.8 &  4.5 & 18.4 \\
23& PSR B$2045-16$	&  3.3 & 3.3 & 2.2 &  5.9 &  2.8 &  6.2 \\
24& PSR B$2310+42$	&  3.2 & 3.2 & 4.0 & 10.3 &  7.2 & 18.5 \\
25& PSR B$2327-20$	&  7.0 & 7.4 & 3.2 & 21.0 & 20.6 & 13.9 \\
\enddata
\end{deluxetable}

\newcommand{\ndg}{\mbox{ $ \nu _{d,g} $ }}
\newcommand{\tdg}{\mbox{ $ \tau _{d,g} $ }}
\begin{deluxetable}{clrrrrr}
\tablenum{6a}
\tablewidth{0pt}
\tablecaption{Decorrelation Bandwidth Measurements from the Literature}
\tablehead
{
Pulsar  & \ndg & C86$^1$ & CWB85 & SW85$^2$ & RA82$^3$ & GRL94$^2$ \\
	& (kHz) & (kHz) & (kHz) & (kHz) & (kHz) & (kHz)  \\ \\
(1) & (2) & (3) & (4) & (5) & (6) & (7) \\
}
\startdata
PSR B$0329+54$ &~~165             &  28     &$\cdots$ &  11     &$\cdots$ & 18      \\
PSR B$0628+28$ &~~203             & 139     &$\cdots$ & 189     & 238     & 45      \\
PSR B$0823+26$ &~~293$^{\dagger}$ & 242     & 91$^a$  & 189     &$\cdots$ & 71      \\
PSR B$0834+06$ &~~454$^{\dagger}$ & 188     & 69$^a$  & 604     & 315     &476      \\
PSR B$0919+06$ &~~256             &  36     & 14$^a$  & 189     &$\cdots$ &$\cdots$ \\
PSR B$1133+16$ &~~816$^{\dagger}$ & 441     &275$^a$  & 491     &$\cdots$ &$\cdots$ \\
PSR B$1237+25$ &~1828             & 451     &230$^a$  &1511     &$\cdots$ &$\cdots$ \\
PSR B$1508+55$ &~~197             &  17     &$\cdots$ & 302     &$\cdots$ &$\cdots$ \\
PSR B$1540-06$ &~~111             & 197     &$\cdots$ &$\cdots$ &$\cdots$ &$\cdots$ \\
PSR B$1604-00$ &~~378             & 375     &343$^b$  &1511     &$\cdots$ &$\cdots$ \\
PSR B$1747-46$ &~~165             &$\cdots$ &$\cdots$ &$\cdots$ &$\cdots$ &$\cdots$ \\
PSR B$1919+21$ &~~285$^{\dagger}$ & 175     & 33$^b$  & 340     & 656     &$\cdots$ \\
PSR B$1929+10$ &~1293             &1009     &946$^b$  & 831     &1439     &$\cdots$ \\
PSR B$2016+28$ &~~206             &  46     & 37$^a$  & 113     &$\cdots$ & 35      \\
PSR B$2020+28$ &~~270             &  99     & 70$^b$  & 227     &$\cdots$ &$\cdots$ \\
PSR B$2045-16$ &~~539             & 366     &$\cdots$ & 604     &$\cdots$ &$\cdots$ \\
PSR B$2310+42$ &~~114             &$\cdots$ &$\cdots$ &$\cdots$ &$\cdots$ &$\cdots$ \\
PSR B$2327-20$ &~~268             &$\cdots$ &$\cdots$ &$\cdots$ &$\cdots$ &$\cdots$ \\
\tablenotetext{\dagger}{Average of measurements of \ndg from different observing sessions}
\tablenotetext{1}{Scaled from values given at 1 GHz.}
\tablenotetext{2}{Scaled from measurements at 408 MHz.}
\tablenotetext{3}{Scaled from measurements at 335 MHz.}
\tablenotetext{a}{Scaled from measurements at 430 MHz.}
\tablenotetext{b}{Scaled from measurements at 320 MHz.}
\tablenotetext{}{} 
\tablenotetext{}{Notes:} 
\tablenotetext{}{\ndg $-$ measurements from present observations}
\tablenotetext{}{C86 $-$ Cordes (1986)}
\tablenotetext{}{CWB85 $-$ Cordes et al. (1985)}
\tablenotetext{}{SW85 $-$ Smith \& Wright (1985)}
\tablenotetext{}{RA82 $-$ Roberts \& Ables (1982)}
\tablenotetext{}{GRL94 $-$ Gupta et al. (1994)}
\enddata
\end{deluxetable}


\begin{deluxetable}{clrcrrr}
\tablenum{6b}
\tablewidth{0pt}
\tablecaption{Scintillation Time Measurements from the Literature}
\tablehead
{
Pulsar  & \tdg & C86$^1$ & CWB85$^@$ & SW85$^2$ & RA82$^3$ & GRL94$^2$ \\
	& (secs) & (secs) & & (secs) & (secs) & (secs) \\ \\
(1) & (2) & (3) & (4) & (5) & (6) & (7) \\
}
\startdata
PSR B$0329+54$ &307             &147      &$\cdots$ &253      &$\cdots$ &149      \\
PSR B$0628+28$ &455             &268      &$\cdots$ &353      &306      &209      \\
PSR B$0823+26$ &126$^{\dagger}$ &173      &$\cdots$ &153      &$\cdots$ & 78      \\
PSR B$0834+06$ &390$^{\dagger}$ &131      &$\cdots$ &284      &262      &272      \\
PSR B$0919+06$ &160             & 63      &$\cdots$ &207      &$\cdots$ &$\cdots$ \\
PSR B$1133+16$ &165$^{\dagger}$ & 77      &$\cdots$ &184      &$\cdots$ &$\cdots$ \\
PSR B$1237+25$ &439             &134      &$\cdots$ &575      &$\cdots$ &$\cdots$ \\
PSR B$1508+55$ &158             & 34      &$\cdots$ &138      &$\cdots$ &$\cdots$ \\
PSR B$1540-06$ &526             &314      &$\cdots$ &$\cdots$ &$\cdots$ &$\cdots$ \\
PSR B$1604-00$ &933             &688      &$\cdots$ &767      &$\cdots$ &$\cdots$ \\
PSR B$1747-46$ &215             &$\cdots$ &$\cdots$ &$\cdots$ &$\cdots$ &$\cdots$ \\
PSR B$1919+21$ &374$^{\dagger}$ &150      &$\cdots$ &245      &272      &$\cdots$ \\
PSR B$1929+10$ &348             &353      &$\cdots$ &376      &379      &$\cdots$ \\
PSR B$2016+28$ &995             &465      &$\cdots$ &583      &$\cdots$ &354      \\
PSR B$2020+28$ &279             &293      &$\cdots$ &245      &$\cdots$ &$\cdots$ \\
PSR B$2045-16$ &138             & 67      &$\cdots$ &123      &$\cdots$ &$\cdots$ \\
PSR B$2310+42$ &309             &$\cdots$ &$\cdots$ &$\cdots$ &$\cdots$ &$\cdots$ \\
PSR B$2327-20$ &432             &$\cdots$ &$\cdots$ &$\cdots$ &$\cdots$ &$\cdots$ \\
\tablenotetext{\dagger}{Average of measurements of \tdg from different observing sessions}
\tablenotetext{@}{Scintillation time measurements are not available.}
\tablenotetext{1}{Scaled from values given at 1 GHz.}
\tablenotetext{2}{Scaled from measurements at 408 MHz.}
\tablenotetext{3}{Scaled from measurements at 335 MHz.}
\tablenotetext{}{} 
\tablenotetext{}{Notes:} 
\tablenotetext{}{\tdg $-$ measurements from present observations}
\tablenotetext{}{C86 $-$ Cordes (1986)}
\tablenotetext{}{CWB85 $-$ Cordes et al. (1985)}
\tablenotetext{}{SW85 $-$ Smith \& Wright (1985)}
\tablenotetext{}{RA82 $-$ Roberts \& Ables (1982)}
\tablenotetext{}{GRL94 $-$ Gupta et al. (1994)}
\enddata
\end{deluxetable}
